\documentclass[12pt]{article}
\usepackage{amsmath}
\usepackage{amssymb}
\numberwithin{equation}{section}

\title{
Recursion relations for Unitary integrals, Combinatorics
 and the Toeplitz Lattice }

\author{
M. Adler\thanks{ Department of Mathematics, Brandeis
University, Waltham, Mass 02454, USA. E-mail:
adler@math.brandeis.edu.  The support of a National
Science Foundation grant \# DMS-01-00782 is
gratefully acknowledged.}~~~~~~ P. van
Moerbeke\thanks{ Department of Mathematics,
Universit\'e de Louvain, 1348 Louvain-la-Neuve,
Belgium and Brandeis University, Waltham, Mass 02454,
USA. E-mail: vanmoerbeke@geom.ucl.ac.be and
@math.brandeis.edu. The support of a National Science
Foundation grant \# DMS-01-00782, a Nato, a FNRS and
a Francqui Foundation grant is gratefully
acknowledged.}}

\date{January 25, 2002}

\catcode`\@=11
\let\c@equation=\relax
\newcounter{equation}[subsection]

\catcode`\@=12

\newcommand{\MAT}[1]{\left(\begin{array}{*#1c}}
\newcommand{\mat}{\end{array}\right)}
\newcommand{\qed}{\leavevmode\unskip\nobreak\penalty200\hskip2pt\null
\nobreak\hfill\rule{1.1ex}{1.1ex}%\parfillskip=0pt
\medbreak
}

\newcommand{\rg}{\rightarrow}

\newcommand{\LR}{{\cal L}}

\newcommand{\VR}{{\cal V}}

\newcommand{\BC}{{\mathbb C}}

\newcommand{\BX}{{\mathbb X}}
\newcommand{\BY}{{\mathbb Y}}
\newcommand{\BV}{{\mathbb V}}
\newcommand{\BZ}{{\mathbb Z}}

\newcommand{\iy}{\infty}
\newcommand{\pl}{\partial}
\newcommand{\al}{\alpha}

\newenvironment{proof}{\medskip\noindent{\it Proof:\/} }{\qed}
\newenvironment
        {remark}{\medskip\noindent\underline{\it Remark:\/} }{\medbreak}

\newcommand{\no}{\nonumber}

\newcommand{\la}{\langle}
\newcommand{\ra}{\rangle}
\newcommand{\ga}{\gamma}

\newcommand{\dt}{\delta}
\newcommand{\Dt}{\Delta}
 \newcommand{\vr}{\varepsilon}

\newcommand{\Lb}{\Lambda}
\newcommand{\tr}{\mbox{tr}}

\newcommand{\BJ}{{\mathbb J}}

\newcommand{\diag}{\operatorname{diag}}

\def\be#1\ee{\begin{equation}#1\end{equation}}
\def\bea#1\eea{\begin{eqnarray}#1\end{eqnarray}}
\def\bean#1\eean{\begin{eqnarray*}#1\end{eqnarray*}}

\newcommand{\Tr}{\operatorname{\rm Tr}}

\newtheorem{definition}{Definition}[section]
\newtheorem{theorem}[definition]{Theorem}
\newtheorem{lemma}[definition]{Lemma}
\newtheorem{corollary}[definition]{Corollary}
\newtheorem{proposition}[definition]{Proposition}

\begin{document}
\maketitle

\tableofcontents

%\newpage

\vspace{1cm}

%\begin{abstract}

%The Toeplitz determinants (of increasing size) associated with the
% symbols $ e^{t(z+z^{-1})} $ or $ \left(1-{\xi}{z} \right)^{\al}
%   \left(1-{\xi}{z^{-1}}  \right)^{\beta}$ satisfy
%   recursion relations, thus expressing all the
%   Toeplitz determinants as a rational function of the first
%   few determinants.
%   A. Borodin found these relations using Riemann-Hilbert methods.
%   The nature of Borodin's relations
%   pointed towards the Toeplitz lattice and its Virasoro algebra,
%   as developed by the authors.
%   In this paper, we take the Toeplitz and Virasoro approach for a
%   fairly large class of symbols, leading to a systematic and
%   simple way of generating such recursion relations.
%    The latter are very naturally expressed
%   in terms of the $L$-matrices appearing in the Lax pair for
%   the Toeplitz lattice equations. As a
%   surprise, we find, compared to Borodin's, a different set of
%   relations, except for the 3-step relations associated with the
%    symbol $ e^{t(z+z^{-1})}$. Moreover, these recursion
% relations define an invariant manifold for the Toeplitz lattice.
% This leads to a ``discrete Painlev\'e property"
% (singularity confinement) for the rational recursion relations,
% as a consequence of the classical ``continuous Painlev\'e
% property" for the Toeplitz lattice.

%\end{abstract}

In a discussion in spring 2001, Alexei Borodin
 showed us recursion
 relations for the Toeplitz determinants going with the
 symbols $ e^{t(z+z^{-1})} $ and $ \left(1-{\xi}{z} \right)^{\al}
   \left(1-{\xi}{z^{-1}}  \right)^{\beta}$. Borodin obtained these
   relations using Riemann-Hilbert
   methods; see the recent work of Borodin \cite{B}
   and also Borodin-Deift \cite{BD} and Baik \cite{Baik}.
   The nature of Borodin's recursion relations
   pointed towards the Toeplitz lattice and its
   Virasoro algebra, introduced by us in \cite{AvM1}.
   In this paper, we take the
   Toeplitz and Virasoro approach for a
   fairly large class of symbols, leading to
   a systematic way of generating recursion relations.
   The latter are very naturally expressed
   in terms of the $L$-matrices appearing in the Toeplitz
   lattice equations. As a
   surprise, we find, compared to Borodin's, a different set of
   relations, except for the 3-step relations associated with the
    symbol $ e^{t(z+z^{-1})}$.

\setcounter{section}{-1}
\section{Introduction and main results}
 The weight %($u_0=0$)
  \be
 \rho(z):=
 e^{P_1(z)+P_2(z^{-1})}z^{\gamma}(1-d_1z)^{\gamma'_1}
  (1-d_2 z)^{\gamma'_2}(1-d_1^{-1}z ^{-1})^{\gamma''_1}
  (1-d_2^{-1}z^{-1})^{\gamma''_2}
 \label{weight}\ee
 with
  \be
  P_1(z):=\sum_1^{N_1} \frac{u_iz^i}{i}
  ~~\mbox{and}
  ~~
  P_2(z^{}):=\sum_1^{N_2} \frac{u_{-i}z^{i}}{i},
  \ee
  has a natural involution
  \be
  \tilde{} : z\leftrightarrow z^{-1},
 \label{involution} \ee
 which induces an involution
 on the following quantities:
   \be
  \tilde{}~ : P_1(z)\leftrightarrow P_2(z^{-1}),~u_i
  \leftrightarrow u_{-i},~
  N_1\leftrightarrow N_2,~\gamma \leftrightarrow
  -\gamma,
   d_i \leftrightarrow d_i^{-1},~
   \gamma_i^{\prime}\leftrightarrow \gamma_i^{\prime\prime}
  .\label{involution2}\ee
  The multiple integral below is known
   to be expressible, both, as the determinant of
   a Toeplitz matrix and as an integral over the group $U(n)$,
     \bea
I_n^{(\vr)}&:=&\frac{1}{n!}
\int_{(S^1)^{n}}|\Dt_n(z)|^{2}
 \prod_{k=1}^n
\left(z_k^{\vr}\rho(z_k)
 \frac{dz_k}{2\pi i z_k}\right)
\no \\
 &=&
 \det\left(\int_{S^1}z^{\vr+i-j}
%e^{\sum_1^{\iy}(t_k z^k-s_kz^{-k})}
 \rho(z)
 \frac{dz}{2\pi i z}\right)_{1\leq i,j\leq n}
 \no\\
 &=&
 \int_{U(n)}  \det \left(U^{\vr}\rho(U) \right) dU
, \label{integral}
 \eea
which for some special choices of $\rho$ has an interesting
interpretation in terms of random permutations; for that matter,
look at the examples in section 4. Consider the basic variables,
with $I_n:=I_n^{(0)}, ~ I_n^{\pm} =I_n^{(\pm 1)}$,
 \be x_n=(-1)^n \frac{I_n^{+}}{I_n} ~~,~~
y_n= (-1)^n \frac{I_n^{-}}{I_n} ~~ \mbox{and} ~~
 v_n=1-x_ny_n=\frac{I_{n-1}I_{n+1}}{I_n^2}.\label{basicvariables}\ee
 See section 1.1 for explanations. Then the basic object
 $$
 I_n=\int_{U(n)}\det \left(\rho(U) \right) dU=
  \det\left(\int_{S^1}z^{i-j}
%e^{\sum_1^{\iy}(t_k z^k-s_kz^{-k})}
 \rho(z)
 \frac{dz}{2\pi i z}\right)_{1\leq i,j\leq n},
 $$
 which appears in several
 problems of random words and permutations,
 is obtained from the $x_n$, $y_n$ and $I_1$,
  by means of the formula
  \be
 I_n=I_1^n \prod_1^{n-1} (1-x_iy_i)^{n-i}.\label{formulaforI}
 \ee

   The following matrices, intimately related to the
Toeplitz lattice, will play an important role in this
work :
 \be
  L_1:=\left(\begin{tabular}{lllll}
$-x_1y_0$  &  $1-x_1y_1$ & ~~ $0$      & ~~ $0$ &   \\
$-x_2y_0$ &  $-x_2y_1$ & $1-x_2y_2$& ~~ $0$   & \\
$-x_3y_0$ &  $-x_3y_1$ & $ -x_3y_2$&  $1-x_3y_3$ & \\
$ -x_4y_0$ &  $ -x_4y_1$ & $-x_4y_2$  & $ -x_4y_3$   &
\\
 & &  &    &  $\ddots$\\
\end{tabular}
\right)
 \label{matrixL}
  \ee
   and
    \be
 L_2:= \left(\begin{tabular}{lllll}
$-x_0y_1$  &  $-x_0y_2$ & $-x_0y_3$     & $-x_0y_4$ &
\\ $1 -x_1y_1$ &  $-x_1y_2$  & $-x_1y_3$& $-x_1y_4$
& \\ ~~$0$       &  $1 -x_2y_2$ & $ -x_2y_3$&
$-x_2y_4$ & \\ ~~$0$       &  ~~$0$      & $ 1
-x_3y_3$  & $ -x_3y_4$   &  \\
 & &  &    &  $\ddots$\\
\end{tabular}
\right).\ee
 The Toeplitz lattice and its relation to the Toda lattice will be
 discussed in Section 1.1.

 Define the matrices, depending on the positive integer
 $n\geq 1$, and the exponents $\gamma,\gamma_i^{\prime}$ and
 $\gamma_i^{\prime\prime}$ in (\ref{weight}),
   \bea
 \LR^{(n)}_1&:=&(aI+b L_1 +c L_1^2)P_1^{\prime}( L_1)
   +c(n+\gamma_1^{\prime}+\gamma_2^{\prime}
   +\gamma) L_1 \nonumber\\
 \LR_2^{(n)}&:=&(cI+b L_2 +aL_2^2 )P_2^{\prime}( L_2)
   +a(n+\gamma_1^{\prime\prime}+\gamma_2^{\prime\prime}
    -\gamma) L_2,
   \label{L-matrices} \eea
  and depending on arbitrary parameters $a,b,c$.
The involution $\tilde{}$, defined in (\ref{involution}) and
(\ref{involution2}) induces involutions
\be
I_n \leftrightarrow I_n, ~I^{+}_n \leftrightarrow
I^{-}_n, x_n  \leftrightarrow  y_n, ~ a\leftrightarrow
c,~b\leftrightarrow b,~\mbox{and so}~~
 L_1  \leftrightarrow L_2^{\top},~~
 \LR^{(n)}_1 \leftrightarrow   \LR^{(n) \top}_2.
 \label{involution1}\ee
Also note that (self-dual case)
 \be
 \rho(z)=\rho(z^{-1}) ~~ \mbox{implies}~~
 x_n=y_n, ~ L_1=L_2^{\top}, ~\LR_1^{(n)}=\LR_2^{(n)\top}
 .\label{selfduality}\ee

 Given a
matrix $A(n)$ containing explicitly the
  parameter $n$,
 the ``{\em discrete derivative}" $\pl_n$ is defined
  as
  \be \pl_n A(n)_{nn}:= A(n+1)_{n+1,n+1}-A(n)_{nn}. \ee

%\newpage

\noindent{\bf Rational relations:} In the
 Theorem 0.1, we show the
 polynomial relationships between consecutive $(x_i,
 y_i)$'s. When the degrees of $P_1$ and $P_2$ differ
 by at most one, they actually lead to inductive rational
 relations, as is stated in Theorem 0.2. These relations are
 obtained by observing that the multiple integral
 (\ref{integral}) satisfy the Toeplitz lattice and
 an SL($2,\BZ$)-set of Virasoro relations in the
 $u_i$-variables; see \cite{AvM1}.

\begin{theorem}

For the weight (\ref{weight}), the vectors %of integrals
$(x_k)_{k\geq 1}$ and $(y_k)_{k\geq 1}$ satisfy two
finite difference relations, involving a finite number
of steps:

 \noindent $\bullet$ {\bf Case 1}. When
$d_1,d_2,d_1-d_2,
   |\gamma_1^{\prime}|+|\gamma_1^{\prime\prime}|,~
   |\gamma_2^{\prime}|+|\gamma_2^{\prime\prime}|
   \neq 0$ in the weight (\ref{weight}), then
  the relations are
 \bea
 \pl_n (\LR_1^{(n)}
   -\LR_2^{(n)})_{n,n}
  ~~~~~+~~~~(c L_1-aL_2)_{nn}~~~~~~~~&=&0
 % \no\\&&
 \label{recurrence1} \\&&\no\\
 \pl_n(v_{n}\LR_1^{(n)}-\LR_2^{(n)}
  )_{n+1,n}
  +\bigl(c L_1^2+b L_1\bigr)_{n+1,n+1}
  -C&=&0, %\no\\&&
  \label{recurrence2}
    \eea
for all $n\geq 1$, and where
 $$
  a=1,~ b=-d_1-d_2,~ c=d_1d_2.$$
$C$ is a constant \underline{independent} of $n$, thus
expressable in terms of the initial value:
 \be
  C:=(v_{1}\LR_1^{(1)}-\LR_2^{(1)}
 )_{2,1}
 +\bigl(c L_1^2+b L_1\bigr)_{1,1}.
 \ee
 Only the first relation is self-dual for the involution
 $\tilde {}$~.

\noindent $\bullet$ {\bf Case 2}. When $d_1\neq
0,\gamma_1^{\prime}
  \neq 0,~
  \gamma^{\prime}_2=
  \gamma^{\prime\prime}_1=
  \gamma^{\prime\prime}_2=d_2=0,
  $
  we may rescale $z$ so that $d_1=-1$ and so
  $$ \rho(z)=z^{\gamma}(1+z)^{\gamma_1^{\prime}}
 e^{P_1(z)+P_2(z^{-1})}
 .$$
 Then the same equations (\ref{recurrence1}) and
 (\ref{recurrence2}) are satisfied, where
  $a,b,c$ can be chosen in two different ways, one
  being the dual of the other, namely $$(a,b,c)=
  (1,1,0)~~\mbox{or}~~(a,b,c)= (0,1,1).$$

\noindent $\bullet$ {\bf Case 3}. $d_1=d_2=
  \gamma^{\prime}_1=
  \gamma^{\prime}_2=
  \gamma^{\prime\prime}_1=
  \gamma^{\prime\prime}_2=0$. Then
  $$
 \rho(z):=
 z^{\gamma} e^{P_1(z)+P_2(z^{-1})} $$
and $a,b,c$ can be chosen totally arbitrary. Here it will be more advantageous to pick different
relations, both of which are polynomials, dual to each
other,
 \bea
 \left\{
 \begin{array}{l}
 \Gamma_n(x,y)%=R^{(3)}_1(x,y)_n
 :=
  \displaystyle{\frac{v_n}{y_n}}
  \left(
  \begin{array}{l}
  -\left( L_1 P_1^{\prime} (
L_1)\right)_{n+1,n+1}
  - \left(  L_2 P_2^{\prime} (
L_2)\right)_{n,n} \\  \\
 + (P_1^{\prime}( L_1))_{n+1,n}
  + (P_2^{\prime}( L_2))_{n,n+1}
 \\
 \end{array}\right)  +nx_n=0
  \\ \\
\tilde\Gamma_n(x,y)%= R^{(3)}_2(x,y)_n
  :=
  \displaystyle{\frac{v_n}{x_n}  }
\left(
\begin{array}{l}
  -\left( L_1 P_1^{\prime} (
L_1)\right)_{n,n}
 -\left(  L_2 P_2^{\prime} (
L_2)\right)_{n+1,n+1} \\  \\
 + (P_1^{\prime}(
L_1))_{n+1,n}
  + (P_2^{\prime}( L_2))_{n,n+1}
  \end{array}\right) +ny_n=0
  \no\\
%\hspace{9cm}-n\frac{x_ny_n}{v_n}=0 . \\
%
\end{array} \right.\\
\label{recurrence3}
   \eea
   \end{theorem}

%   \newpage

\begin{theorem}

Requiring $~N_1=N_2\mbox{ or } N_2\pm 1$ in the weight
 (\ref{weight}), the
$x_n$ and $y_n$'s can be expressed rationally in terms of lower
$x$'s and $y$'s, and thus, from (\ref{formulaforI}), $I_n$ can
 be expressed in the terms of the $x$'s and $y$'s.
 To be precise,

 \noindent $\bullet$ {\bf
Case 1} leads to two inductive
  \underline{\em rational} $N_1+N_2+4$-step relations
    \bean
    x_n &=& F_n( x_{n-1},y_{n-1},\ldots,x_{n-N_1-N_2-3},
    y_{n-N_1-N_2-3}) \\
    y_n &=& G_n( x_{n-1},y_{n-1},\ldots,x_{n-N_1-N_2-3},
    y_{n-N_1-N_2-3}).
   \eean

\noindent $\bullet$ {\bf Case 2} leads to two
inductive
  \underline{\em rational} $N_1+N_2+3$-step relations
  (\ref{recurrence1}) and (\ref{recurrence2}),
  such that\footnote{Both solutions can be used, when
  $N_1=N_2$.}
  $$
  \begin{array}{ll}
   %~~ \mbox{or}~~
     \mbox{when $N_1=N_2$ or $N_1=N_2+1$},&
     \mbox{use} ~~(a,b,c)= (1,1,0)\\
    \mbox{when $N_2=N_1$ or $N_2=N_1+1$},&
     \mbox{use}~~(a,b,c)= (0,1,1)
   .\end{array}
   $$
 Thus, we find rational functions $F_n$ and $G_n$:
    \bean
    x_n &=& F_n( x_{n-1},y_{n-1},\ldots,x_{n-N_1-N_2-2},
    y_{n-N_1-N_2-2}) \\
    y_n &=& G_n( x_{n-1},y_{n-1},\ldots,x_{n-N_1-N_2-2},
    y_{n-N_1-N_2-2}).
   \eean

\noindent $\bullet$ {\bf Case 3} leads to two
inductive $N_1+N_2+1$-step
  \underline{\em rational} relations
    \bean
    x_n &=& F_n( x_{n-1},y_{n-1},\ldots,x_{n-N_1-N_2},
    y_{n-N_1-N_2}) \\
    y_n &=& G_n( x_{n-1},y_{n-1},\ldots,x_{n-N_1-N_2},
    y_{n-N_1-N_2}).
   \eean

\end{theorem}

%\newpage

   \begin{corollary}

   For the self-dual weight
 $$\rho(z)
  =
  e^{\sum_1^N \frac{u_i}{i}(z^i+z^{-i})},
  $$
  the polynomial\footnote{The matrix
 $L_1$ appearing in (\ref{recurrence3self}) is the matrix (\ref{matrixL}), with
 $y_i=x_i$.} in
$x_{k-N},x_{k-N+1},\ldots,x_{k},$ $\ldots,x_{k+N}$,
  {\footnotesize\bea
  \Gamma_k&:=&  kx_k- \frac{v_k}{x_k}\left(\left(\sum_1^N
u_iL_1^i\right)_{k+1,k+1}+ \left(\sum_1^N
u_iL_1^i\right)_{k,k}-2\left(\sum_1^N
u_iL_1^{i-1}\right)_{k+1,k}\right)\no\\
&=&  kx_k +v_k \sum_1^N u_i
 \left( \sum_{j=k-1}^{k+i-1}x_{j+1}(L_1^{i-1})_{k+1,j+1}+
   \sum_{j=k-i+1}^{k+1}x_{j-1}(L_1^{i-1})_{jk} \right) =0
\no\\ \label{recurrence3self}\eea
  }
 leads to recurrence relations
  $$
   x_n = F_n( x_{n-1}, \ldots,x_{n-2N}).
    $$

  \end{corollary}

   \remark In the self-dual case, i.e., when $\rho (z) =
\rho(z^{-1})$, the first equation (\ref{recurrence1}) vanishes
identically and the two equations in (\ref{recurrence3}) become
identical. Only one equation is required, since all $x_n=y_n$.

\remark
 By the duality $a\leftrightarrow
c,~b\leftrightarrow
 b,~ L_1 \leftrightarrow L_2^{\top}$, equations
 (\ref{recurrence1}) and (\ref{recurrence2}) map into

  \be
 \left\{\begin{array}{l} \pl_n (\LR_1^{(n)}-\LR_2^{(n)})_{n,n}
 ~~~~~+~~~~~(c L_1-aL_2)_{nn}~=~0  \\ \\
 \pl_n(\LR_1^{(n)}-v_{n}\LR_2^{(n)}
 )_{n,n+1}
  -\bigl(aL_2^2+b  L_2\bigr)_{n+1,n+1}
 =C',\end{array}\right. \ee
 where $C'$ is now:
 \be
  C':=(\LR_1^{(1)}-v_{1}\LR_2^{(1)}
 )_{1,2}
 -\bigl(a  L_2^2+b L_2\bigr)_{1,1}.
 \ee

%\newpage

\noindent{\bf An invariant manifold:} The relations
 appearing in Theorem 0.1 for each of the cases
 happen to define an invariant manifold
 for the first Toeplitz flow. We shall do this here
 for case 3, where
   $$
 \rho(z):=
 z^{\gamma} e^{P_1(z)+P_2(z^{-1})}. $$
  This case has the extra feature
 that the relations themselves (\ref{recurrence3})
 satisfy an interesting system of
 differential equations. These
 statements will be established in section 3, as an
 immediate consequence
 of the Virasoro relations satisfied by the multiple integrals.

\begin{theorem} Let $x_n=x_n(t_1,s_1)$ and $y_n
=y_n(t_1,s_1)$ flow according to the differential
equations ($v_n:=1-x_ny_n$)
 \bea
  \frac{\pl x_k}{\pl t_1}=v_kx_{k+1}  &~~~~~&
\frac{\pl y_k}{\pl t_1}=-v_ky_{k-1} \no \\&&
 \hspace{3cm}(\mbox{\bf Toeplitz
Lattice}) \label{ToeplitzFlow}\\ \frac{\pl x_k}{\pl s_1}=v_kx_{k-1}
&~~~~~& \frac{\pl y_k}{\pl s_1}=-v_ky_{k+1}  .
 \no\\\no
 \eea
and the $u_i$, appearing in the polynomials $P_1(z)$
and $P_2(z)$, according to
  $$\frac{\pl u_k}{\pl t_1}=\dt_{k,1}
  ~~~~~~~~~~~~\frac{\pl u_k}{\pl s_1}=-\dt_{k,-1}.$$
 Then:

 \noindent
  {\bf(i)} The polynomial recurrence relations $ \Gamma_n$ and
 $\tilde\Gamma_n$, defined in (\ref{recurrence3}), satisfy
 the differential equations
\bea
 \frac{\pl}{\pl \left\{{t_1 \atop s_1}\right\}}\Gamma_n
 &=&
  v_n\Gamma_{n\pm 1}+ x_{n\pm 1}
   \left(x_n \tilde\Gamma_n   -
   y_n \Gamma_n\right)\no\\&&\no\\
   \frac{\pl}{\pl \left\{{t_1 \atop s_1}\right\}}
    \tilde\Gamma_n
 &=&
  -v_n\tilde\Gamma_{n\mp 1} +  y_{n\mp 1}
   \left(x_n \tilde\Gamma_n  -
   y_n \Gamma_n\right).
 \label{gamma-ode1}\eea
\noindent
  {\bf(ii)} The locus % ${\frak M}^{(1)}$ and
  ${\frak M} $
 is an invariant manifold for the $t_1$ and $s_1$-flows
(\ref{ToeplitzFlow}) above, where
   \be
    {\frak
M}%^{(k)}
 :=\bigcap_{n\geq 0} \left \{ (x_k,y_k)_{k\geq
0}~,~~\mbox{such that}~~
\Gamma_n(x,y)=0%R^{(k)}_1(x,y)_n=0,~R^{(k)}_2(x,y)_n=0
 \mbox{ and }\tilde\Gamma_n(x,y)=0\right\}
. \label{M-locus}
  \ee

\end{theorem}

\begin{corollary} Let $x_n=x_n(t)$ flow according to the differential
equations ($v_n:=1-x_n^2$)
\be
 \frac{\pl x_n}{\pl
t}=v_n(x_{n+1}-x_{n-1}),
 \label{SymToeplitzFlow}
 \ee
 which is obtained by taking the linear combination
  $\frac{\pl}{\pl t_1}- \frac{\pl}{\pl s_1}$ of the
  Toeplitz vector fields above and setting all $x_k=y_k$.
Let the $u_i$, appearing in the self-dual weight
 $$\rho(z)
  =
  e^{\sum_1^N \frac{u_i}{i}(z^i+z^{-i})},
  $$
 flow according to
  $$\frac{\pl u_k}{\pl t}=\dt_{k,1}
   .$$
 Then:

 \noindent
  {\bf(i)} The polynomial recurrence relations $ \Gamma_n$
  , as in (\ref{recurrence3self}), satisfy
 the differential equations
 \be
 \frac{\pl \Gamma_n}{\pl t}=v_n (\Gamma_{n+1}-
 \Gamma_{n-1}) .\label{Gamma-ode-intro}
 \ee

\noindent
  {\bf(ii)} The locus % ${\frak M}^{(1)}$ and
  ${\frak N} $
 is an invariant manifold for the $t$-flow
 (\ref{SymToeplitzFlow})
above, where
   \be
    {\frak
N}%^{(k)}
 :=\bigcap_{n\geq 0} \left \{ (x_k)_{k\geq
0}~,~~\mbox{such that}~~
\Gamma_n(x,x)=0%R^{(k)}_1(x,y)_n=0,~R^{(k)}_2(x,y)_n=0
  \right\}
. \label{N-locus}
  \ee

\end{corollary}

%\newpage

\noindent{\bf Singularity confinement:}
 For the self-dual weight
 $$\rho(z)
  =
  e^{\sum_1^N \frac{u^i}{i}(z^i+z^{-i})},
  $$
 the polynomial relations
 (remember Corollary 0.3) in $x_{k-N},x_{k-N+1},\ldots,x_{k},
  $ $ \ldots,x_{k+N}$,
 {\footnotesize \bean
   %\Gamma_k
   0&=&kx_k- \frac{v_k}{x_k}\left(\left(\sum_1^N
u_iL_1^i\right)_{k+1,k+1}+ \left(\sum_1^N
u_iL_1^i\right)_{k,k}-2\left(\sum_1^N
u_iL_1^{i-1}\right)_{k+1,k}\right)%\no\\&&+kx_k
 %\label{recurrence3self}
  \eean
  }
 lead, in effect, to rational recurrence relations in the $x_i$,
 \be
  x_k= F_k (x_{k-1},\ldots, x_{k-2N}; u_1,\ldots,u_N)
,\label{recurrence4}
 \ee
 depending rationally on the coefficients
 $u_1,\ldots, u_N$ appearing in the weight $\rho(z)$.

%\newpage

 The following Theorem tells us -roughly speaking- that the recurrence
 relations (\ref{recurrence4}) for special initial
 condition leads to a solution, where one $x_n$
 blows up and all
 other $x_k$ are finite. This is a kind of {\em discrete
 Painlev\'e property}, called ``singularity
 confinement"; see Grammaticos, Nijhoff and Ramani
\cite{Nijhoff}, who define this to be discrete Painlev\'e
recursion relations. For these recurrence equations
 (\ref{recurrence4}),
 the precise analytical statement
 of this phenomenon is stated in
 Corollary 0.7, claiming there is a {\em generic}
 solution with the kind
 of singularity above. The
 technique used here to prove Corollary 0.7 is to
 deform the variables $x_k$ and $y_k$ by means of the
 Toeplitz lattice; part {\bf (i)} of Theorem 0.6 below
 shows that the Toeplitz lattice has a generic
 solution $x_0,x_1,\ldots$, with all $x_k,~ k\neq n$ finite and
 one $x_n$ blowing up. This is reminiscent of the
 Painlev\'e property of {\em algebraic integrable systems},
 which originates in the work of S. Kowalewski; see \cite{AvM0}
  and references within. Part {\bf (ii)} of Theorem 0.6
 shows that these series can be made
 to stay within the locus $\frak N$, by restricting
 the free parameters. The proof of Theorem 0.6 and
 Corollary 0.7, which will be given in a subsequent paper,
 uses heavily the ideas of Theorem 0.4 and Corollary 0,5.

\begin{theorem}

{\bf (i)} Consider the system of differential equations (with boundary
condition $x_0=1$)
  \be \frac{\pl x_k}{\pl
t}=(1-x_k^2)(x_{k+1}-x_{k-1}), \mbox{ for } k=0, 1,
2,\ldots,
 \label{ode}\ee
and for a fixed, but arbitrary integer $n>0$, let
 \be
\ldots, \al_{n-3} ,\al_{n-2},c,d ,\al_{n+2} ,\al_{n+3}, \ldots
\label{free}
 \ee
 be free parameters. Then the system (\ref{ode})
has a unique ``formal" Laurent solution,
 with $x_n$ and only $x_n$ blowing up, having the
 form:
%\footnote{$v_k^{(0)}:=1-\al_k^2.$}
\be\begin{array}{l}
  x_k(t)=\al_k+\ldots
   ,~~~        \mbox{ for } |k-n|\geq 2\\ \\
  x_{n-1} (t)=\pm1+ct+\ldots  \\ \\
  x_n(t)= \frac{1}{t}
     (\mp\frac{1}{2}+\frac{c-d}{8} t
                   +\ldots)\\ \\
  x_{n+1}(t)=\mp 1+dt+\ldots~.
\end{array}
\label{series}
\ee
 The coefficients in the series (\ref{series}) are
polynomials in the free parameters (\ref{free}).
 This solution is generic,
 since $$\#\{\mbox{free parameters}\}+1=
  \# \{\mbox{variables}\},$$
 with the ``$1$" accounting for the $t$-parameter.

\bigbreak

 \noindent{\bf (ii)} Given the $2N-1$ free parameters
 $\al_{n-2N},\ldots, \al_{n-2}$, the series
 (\ref{series}) above are ``formal" Laurent
 solutions to the recurrence relations
 (\ref{recurrence3self})
\be
  x_k= F_k (x_{k-1},\ldots, x_{k-2N}; u_1+t,\ldots,u_N)
,\label{recurrence}
\ee
with the remaining free parameters
 $c,d,\al_i~~\mbox{for}~~i\leq n-2N-1~~\mbox{or}~~i\geq n-2
% \al_{n+2}, \al_{n-2},\ldots
,$
being rational functions of
 $\al_{n-2N},\ldots, \al_{n-2}$ and the parameters
 $u=(u_1,\ldots, u_N)$.

 \end{theorem}

This Theorem leads to the ``Painlev\'e singularity confinement"
property for the recursive equations (\ref{recurrence4});  the
precise statement goes as follows:

\begin{corollary} ({\bf Singularity confinement})
  Given arbitrary initial data
 $$
 (x_{n-2N},\ldots, x_{n-2})=
  (x^{(0)}_{n-2N},\ldots, x^{(0)}_{n-2})=:\gamma
 $$
  and setting
 \be
   x_{n-1}=\pm 1+\vr,\label{EpsilonInitialCondition}
 \ee
 the recurrence relations (\ref{recurrence4}), namely
 $$
  x_k= F_k (x_{k-1},\ldots, x_{k-2N}; u_1,\ldots,u_N)
,\label{recurrence}
 $$
  have a ``{\em generic}" formal series solution in $\vr$ of the
 form (i.e., depending on $2N-1$ degrees of freedom)
 \be
 \begin{array}{l}
  x_{n-1} (t)=\pm 1+\vr  \\ \\
  x_n(t)= \displaystyle{\frac{1}{\vr}}
     \left(x_n^{(0)}(\ga,u)+O(\vr)\right)\\ \\
  x_{n+1}(t)=\mp 1+O(\vr) \\ \\
  x_k(t)=x^{(0)}_k(\ga,u)+O(\vr)
   ,~~~        \mbox{ for } k\geq n+ 2,
\end{array}
%\label{solution1}
 \ee
with all coefficients of the $\vr$-series depending rationally on
$\ga=(x^{(0)}_{n-2N},\ldots, x^{(0)}_{n-2})$ and $u:=(u_1,\ldots,
u_N)$.

\end{corollary}

\remark The initial condition (\ref{EpsilonInitialCondition}) is
the most general initial condition leading to blow-up at the $n$th
step.

\medbreak

\noindent Theorem 0.6 and Corollary 0.7 will be established
elsewhere, as well as analogous statements that can be made for
the non-symmetric weight
  $$\rho(z)
  =
  e^{\sum_1^N (\frac{u_i}{i}z^i+\frac{u_{-i}}{i}z^{-i})
  }.
  $$

\noindent{\bf Examples: } Several examples will be discusssed in
section 4. It is also interesting to point out that each of the
examples discussed in that section are related to random
permutations, random words and point processes. They also admit a
representation as a Fredholm determinant of an interesting kernel;
concerning the latter, see Borodin and Okounkov \cite{BO}.

%\vspace{1cm}

%\newpage

\section{The Toeplitz lattice and its Virasoro algebra}
\subsection{The Toeplitz lattice}
Consider the inner-product on the circle
 \be
 \la f(z),g(z)\ra_{t,s}:=\oint_{S^{1}} \frac{dz}{2\pi i z}f(z)g(z^{-1})
 e^{\sum_1^{\iy}(t_iz^i-s_iz^{-i})},
 \ee
  the associated moments
  $\mu_{k-\ell}(t,s):=\la y^k, z^{\ell}\ra_{t,s}$, and
  the determinants ($\tau$-functions)
  \bea \tau_n(t,s)
  &:=& \det \left( \mu_{k-\ell }(t,s)\right)_{0\leq k,\ell
\leq n-1}\nonumber\\
  &=& \frac{1}{n!}
\int_{(S^1)^{n}}|\Dt_n(z)|^{2}
 \prod_{k=1}^n
\left(e^{\sum_1^{\iy}(t_i z_k^i-s_iz_k^{-i})}
 \frac{dz_k}{2\pi i z_k}\right)\nonumber\\&&\nonumber\\
&=&\int_{U(n)}e^{\sum_1^{\iy}\Tr (t_iM^i-s_i\bar
M^i)}dM
 \eea
  and
 \bea \tau^{\pm}_n(t,s)&=&
  \det \left( \mu_{k-\ell\pm 1
}(t,s)\right)_{0\leq k,\ell \leq n-1}, \nonumber
\\
 &=&
\frac{1}{n!} \int_{(S^1)^{n}}|\Dt_n(z)|^{2}
 \prod_{k=1}^n
\left(z_k^{\pm 1} ~e^{\sum_1^{\iy}(t_i
z_k^i-s_iz_k^{-i})}
 \frac{dz_k}{2\pi i z_k}\right)\nonumber\\&&\nonumber\\
&=&\int_{U(n)}(\det M)^{\pm 1}e^{\sum_1^{\iy}\Tr
(t_iM^i-s_i\bar M^i)}dM
 \eea
 The following $\tau$-function expressions are actually
polynomials
\footnote{For $\al\in \BC$, define
    $[\al]:=(\al,\al^2/2,\al^3/3,\dots)\in \BC^{\iy}$.}
     \bea
 p_n^{(1)}(t,s;u)&=&
  u^n\frac{\tau_n(t-[u^{-1}],s)}{\tau_n(t,s)}
   \nonumber\\&=&
 \frac{1}{n!\tau_n}
\int_{(S^1)^{n}}|\Dt_n(z)|^{2}
 \prod_{k=1}^n
\left((u-z_k) e^{\sum_1^{\iy}(t_i z_k^i-s_iz_k^{-i})}
 \frac{dz_k}{2\pi i z_k}\right)
   \nonumber\\&=&
    \frac{1}{\tau_n(t,s)}
  \int_{U(n)}\det (uI-M)e^{\sum_1^{\iy}
  \Tr (t_iM^i-s_i\bar M^i)}dM\nonumber\\
\eea
\bea  p_n^{(2)}(t,s;u)&=&
   u^n\frac{\tau_n(t,s+[u^{-1}])}{\tau_n(t,s)}
 \nonumber\\&=&
 \frac{1}{n!\tau_n}
\int_{(S^1)^{n}}|\Dt_n(z)|^{2}
 \prod_{k=1}^n
\left((u-z^{-1}_k) e^{\sum_1^{\iy}(t_i
z_k^i-s_iz_k^{-i})}
 \frac{dz_k}{2\pi i z_k}\right)
   \nonumber\\&=&
   \frac{1}{\tau_n(t,s)}
  \int_{U(n)}\det (uI-\bar M)e^{\sum_1^{\iy}
  \Tr (t_iM^i-s_i\bar M^i)}dM
  \eea
 and are bi-orthogonal for the inner-product above
\be  \la p_n^{(1)},p_m^{(2)}\ra_{t,s}=\delta_{nm}
h_n(t,s),~~\mbox{  with  }
 h_n:= \frac{\tau_{n+1}}{\tau_n},
  \ee
Define\footnote{
    $\tilde\pl=(\pl/\pl t_1,(1/2)\pl/\pl t_2,(1/3)\pl/
     \pl t_3,\dots)$,
    and $p_k$ are the elementary Schur functions:
    $\sum_{k=0}^\infty p_k(t)z^k:=\exp(\sum_{i=1}^\infty t_iz^i)$.
}
  \bea
 x_n (t,s)
&:=&p_n^{(1)}(t,s;0)\no\\
 &=&
 \frac{(-1)^{n}}{n!\tau_n}
\int_{(S^1)^{n}}|\Dt_n(z)|^{2}
 \prod_{k=1}^n
\left(z_k  e^{\sum_1^{\iy}(t_i z_k^i-s_iz_k^{-i})}
 \frac{dz_k}{2\pi i z_k}\right)
 \no  \\
 &=&\frac{p_n(-\tilde
\pl_t)\tau_n(t,s)}{\tau_n(t,s)}
=(-1)^n \frac{\tau^+_n(t,s)}{\tau_n(t,s)}
\nonumber\\
 y_n(t,s):
 &:=&p_n^{(2)}(t,s;0)
 \no\\
 &=&
 \frac{(-1)^{n}}{n!\tau_n}
\int_{(S^1)^{n}}|\Dt_n(z)|^{2}
 \prod_{k=1}^n
\left(z_k^{-1} e^{\sum_1^{\iy}(t_i z_k^i-s_iz_k^{-i})}
 \frac{dz_k}{2\pi i z_k}\right) \no\\
 &=&\frac{p_n(\tilde \pl_s)
  \tau_n(t,s)}{\tau_n(t,s)}
  =(-1)^n \frac{\tau^-_n(t,s)}{\tau_n(t,s)}
 .
 \label{x,y} \eea
\noindent Throughout the paper, set\footnote{By
computing $\la
p_{n+1}^{(1)}(u)-up_{n}^{(1)}(u),p^{(2)}_{m+1}(u)
 -up^{(2)}_{m}(u)\ra$ in two different ways, in a straightforward way and
 in another way, using
  \bean
p^{(1)}_{n+1}(u)-up_n^{(1)}(u)&=&p^{(1)}_{n+1}(0)u^np_n^{(2)}(u^{-1})\nonumber\\
p^{(2)}_{n+1}(u)-up_n^{(2)}(u)&=&p^{(2)}_{n+1}(0)u^np_n^{(1)}(u^{-1}).
\eean}
  \bea
 v_n:=  1-x_ny_n
    &=&
      1-p_n^{(1)}(t,s;0)p_n^{(2)}(t,s;0) \no\\
    &=&
      \frac{h_{n}}{h_{n-1}}=
      \frac{\tau_{n+1}\tau_{n-1}}{\tau_{n}^2}
  \eea
 In \cite{AvM1}, it was pointed out that the quantities $x_n$ and $y_n$ satisfy the following integrable
Hamiltonian system
 \bea
 \frac{\pl x_n}{\pl
t_i}=(1-x_ny_n)\frac{\pl H^{(1)}_i}{\pl y_n}  &~~~~~&
\frac{\pl y_n}{\pl t_i}=-(1-x_ny_n)\frac{\pl
H^{(1)}_i}{\pl x_n} \nonumber \\ \frac{\pl x_n}{\pl
s_i}=(1-x_ny_n)\frac{\pl H^{(2)}_i}{\pl y_n}  &~~~~~&
\frac{\pl y_n}{\pl s_i}=-(1-x_ny_n)\frac{\pl
H^{(2)}_i}{\pl x_n}
  , \label{Toeplitz}\\ &&\hspace{2.5cm}(\mbox{\bf Toeplitz lattice})
 \nonumber \eea
 with initial condition $x_n(0,0)=y_n(0,0)=0$ for
$n\geq 1$ and boundary condition $x_0(t,s)=y_0(t,s)=1$. This fact
will be established in Proposition 1.1 below. The traces
 \be H^{(k)}_i=-
\frac{1}{i}\Tr~{ L} _k^i,~~i=1,2,3,...,~~k=1,2
 \label{Hamiltonians}
 \ee of
the matrices ${ L}_i$ below are integrals in
involution with regard to the symplectic structure
 $$ \omega := \sum_1^{\iy}
\frac{dx_k \wedge dy_k}{1-x_ky_k}, $$
  where $L_1$ and $L_2$ are given by the `` rank 2" semi-infinite
matrices $$L_1 := \left(\begin{tabular}{lllll}
$-x_1y_0$  &  $1-x_1y_1$ & ~~ $0$      & ~~ $0$ &   \\
$-x_2y_0$ &  $-x_2y_1$ & $1-x_2y_2$& ~~ $0$   & \\
$-x_3y_0$ &  $-x_3y_1$ & $ -x_3y_2$&  $1-x_3y_3$ & \\
$ -x_4y_0$ &  $ -x_4y_1$ & $-x_4y_2$  & $ -x_4y_3$   &
\\
 & &  &    &  $\ddots$\\
\end{tabular}
\right)
 $$ and $$L_2:= \left(\begin{tabular}{lllll}
$-x_0y_1$  &  $-x_0y_2$ & $-x_0y_3$     & $-x_0y_4$ &
\\ $1 -x_1y_1$ &  $-x_1y_2$  & $-x_1y_3$& $-x_1y_4$
& \\ ~~$0$       &  $1 -x_2y_2$ & $ -x_2y_3$&
$-x_2y_4$ & \\ ~~$0$       &  ~~$0$      & $ 1
-x_3y_3$  & $ -x_3y_4$   &  \\
 & &  &    &  $\ddots$\\
\end{tabular}
\right).$$
Written out, the differential equations (\ref{Toeplitz})
read\footnote{Introduce the inner-product
 $$ \la A,B
\ra=\tr AB^{\top} ,$$
 which differentiated behaves as
 $$ \frac{\pl}{\pl x}~ \tr A^n =
  \la  n A^{n-1}, \frac{\pl A}{\pl x}  \ra
   .$$}
\begin{eqnarray}
\frac{\pl x_n}{\pl t_i} &=& (1-x_n y_n) \frac{\pl H^{(1)}_i}{\pl
y_n}
 \no\\
 &=&
  \frac{h_n}{h_{n-1}} \frac{\pl}{\pl y_n} \left(-\frac{1}{i} Tr
L^i_1\right), \mbox{ using } 1-x_n y_n=\frac{h_n}{h_{n-1}}
 \no\\
&=&
 \frac{h_n}{h_{n-1}} \left\langle L_1^{i-1}, -\frac{\pl L_1}{\pl
y_n}\right\rangle
 \no\\
&=&
 \frac{h_n}{h_{n-1}} \left\langle L_1^{i-1},
\begin{array}{c}
\hspace{11mm}\stackrel{n+1}{\downarrow}\\
\MAT{8}
 & & & & & 0 \\
 && & & &\vdots& & \\
 O& & & & &0& &O\\
& & & & &x_n &&\\
 & & & & &x_{n+1}\\
 & & & & &x_{n+2}& &\\
&& & & &\vdots \mat
\end{array}\leftarrow n
 \right\rangle\no\\
&=&\frac{h_n}{h_{n-1}} \sum_{j\geq n} (L_1^{i-1 })_{n+1,j} x_j ,
\label{x-t-flow}
 \end{eqnarray}

and similarly

\begin{eqnarray}
\frac{\pl x_n}{\pl s_i} &=& (1-x_n y_n) \frac{\pl H^{(2)}_i}{\pl
y_n}
 \no\\
% &=&
%  \frac{h_n}{h_{n-1}} \frac{\pl}{\pl y_n} \left(-\frac{1}{i} Tr
%L^i_2\right), \mbox{ using } 1-x_n y_n=\frac{h_n}{h_{n-1}}
% \no\\
 &=&
  \frac{h_n}{h_{n-1}} \left\langle L_2^{i-1}, -\frac{\pl L_2}{\pl
y_n}\right\rangle
\no \\
&=&
 \frac{h_n}{h_{n-1}} \left\langle L_2^{i-1},\begin{array}{c}
\hspace{0.8cm}\stackrel{n}{\downarrow}\\
\begin{array}{c}
\\
%=
\\
\\
\\
 \\
\end{array}
\MAT{8}
  & &  & &x_0& & & \\
 & & & &\vdots& && \\
 & O & & &x_{n}& O\\
& &  & &0&&& \\
 &  & & &0\\
 & & & &\vdots& &
\mat\end{array}
 \right\rangle \no\\
&=&\frac{h_n}{h_{n-1}} \sum^{n+1}_{j=1} (L_2^{i-1 })_{n,j}
x_{j-1}\no\\
&=&\frac{h_n}{h_{n-1}} \sum^{n+1}_{j=n-i+1} (L_2^{i-1 })_{n,j}
x_{j-1}~,
 \label{x-s-flow}
 \end{eqnarray}
using in the last identity the obvious fact that
  \be
 (L_2^{\al})_{ij} =0, \mbox{   unless} ~~j\geq i-\al.
  \label{rel1}
  \ee
By the duality, mentioned in (\ref{invol}) below, one reads off
the differential equations for the $y_n$'s.

\noindent Setting
$$ \hat L_1:= hL_1h^{-1}~~~    \mbox{and}~~~
 \hat L_2:= L_2, $$
  the ``rank
2"-structure of $\hat L_1$ and $\hat L_2$ is preserved
by the equations
 \bea &&\hspace{-.5cm} \frac{\pl
\hat L_i}{\pl t_n}=\bigl[\bigl(\hat L_1^n\bigr)_+,\hat
L_i\bigr] \quad\hbox{and}\quad \frac{\pl \hat L_i}{\pl
s_n}=\bigl[\bigl(\hat L_2^n\bigr)_-,\hat L_i\bigr]
\quad i=1,2~\mbox{and}~n=1,2,\dots \no\\
&&
\mbox {\bf
\hspace{7cm}(Two-Toda Lattice)}\label{Toda}
 \eea

\noindent The Toda and toeplitz lattices have an involution,
compatible with the involution $\tilde{}$ , introduced in
(\ref{involution}),
 \be
 x_n \leftrightarrow y_n,~~t_n\leftrightarrow -s_n,~~
  L_1\leftrightarrow L_2^{\top}.
  \label{invol}\ee

The Proposition below was merely taken for granted in \cite{AvM1}.
Here we give a complete proof.

\begin{proposition} The two-Toda lattice flows (\ref{Toda}) are equivalent to
the Hamiltonian Toeplitz lattice flows (\ref{Toeplitz}).
\end{proposition}

\proof From the general 2-Toda theory, as related to biorthogonal
polynomials, we know that the vector (see \cite{AvM2})
 \be
\Psi_1:=(\Psi_{1,n})_{n\geq0}=e^{\sum_1^{\iy} t_iz^i}
\left(p_n^{(1)} (t,s;z)\right)_{n\geq 0}
 \label{Baker}
 \ee
 with
 \be
\Psi_{1,n}=e^{\sum_1^{\iy} t_iz^i} z^n
\frac{\tau_n(t-[z^{-1}],s)}{\tau_n(t,s)}= e^{\sum_1^{\iy} t_iz^i}
p_n^{(1)} (t,s;z),
 \label{Baker1}
 \ee
  is an eigenvector for the matrix $L_1$,
 $$
L_1\Psi_1=z\Psi_1,
 $$
and (\ref{Baker}) satisfies the differential equations
 \be
\frac{\pl \Psi_1}{\pl t_n}=(\hat L_1^n)_+ \Psi_1 \mbox{ and
}\frac{\pl \Psi_1}{\pl s_n}=(\hat L_2^n)_- \Psi_1.
 \label{Bakerode}
 \ee

So, from (\ref{Baker}), (\ref{Baker1}) and (\ref{Bakerode}), it
follows that the vector
 $p^{(1)} (z):=(p^{(1)}_n (t,s;z))_{n\geq 0}$
 satisfies the differential equations
\begin{eqnarray}
\frac{\pl p^{(1)} (z)}{\pl t_n}&=&(\hat L_1^n)_+ p^{(1)} (z)-z^n
p^{(1)}
(z)\no\\
\frac{\pl p^{(1)} (z)}{\pl s_n}&=&(\hat L_2^n)_- p^{(1)}
(z).\label{polode}
\end{eqnarray}
Since (\ref{x,y}) implies
$p^{(1)}(t,s;0)=(x_0,x_1,\ldots)^{\top}$, the differential
equations (\ref{polode}) evaluated at $z=0$ read
$$
\frac{\pl x}{\pl t_i}=(\hat L_1^i)_+ x ~\mbox{ and }~\frac{\pl
x}{\pl s_i}=(\hat L_2^i)_- x,
$$
yielding componentwise (remember $\hat L_1=hL_1h^{-1}$)
  \bea
\frac{\pl x_n}{\pl t_i}
 &=&h_n\sum_{j\geq n} (L^i_1)_ {n+1,j+1}
\frac{x_j}{h_j}\\
\frac{\pl x_n}{\pl s_i}
 &=&
  \sum_{1\leq j\leq n} (L^i_2)_ {n+1,j} x_{j-1}
 =
  \sum_{j=n+1-i}^n (L^i_2)_ {n+1,j} x_{j-1}.
 \label{x-Lax-flow}\eea
 The point of Proposition 1.1 is to show
that the equations (\ref{x-Lax-flow}), obtained via the 2-Toda
lattice, are equivalent to the equations (\ref{x-t-flow}),
(\ref{x-s-flow}), coming from the Toeplitz lattice, i.e., we must
show, for $n\geq 0$,

\bea
 \sum_{j\geq n} (L_1^{i})_{n+1,j+1} \frac{x_{j}}{h_j}
 &=&
  \frac{1}{h_{n-1}}
   \sum_{j\geq n} (L_1^{i-1})_{n+1,j} x_{j}
   \label{Ham-Lax1}\\
 \sum_{j=n-i+1}^{n} (L_2^{i})_{n+1,j} x_{j-1}
 &=&
  \frac{h_n}{h_{n-1}}
   \sum_{j=n-i+1}^{n+1} (L_2^{i-1})_{nj} x_{j-1}
   \label{Ham-Lax2}\eea
by duality, it suffices to show equations just for the
$x_n$-variables.

To show (\ref{Ham-Lax1}), compute, using $x_jy_j =1- h_j/h_{j-1}$:
\begin{eqnarray*}
\sum_{j\geq n} (L^i_1)_{n+1, j+1} \frac{x_j}{h_j}
 &=&
  \sum_{r\geq j\geq n} (L^{i-1}_1)_{n+1,r} (L_1)_{r,j+1}
\frac{x_j}{h_j}\\
&=&\sum_{r\geq j\geq n} (L^{i-1}_1)_{n+1, r} (\dt_{r,j}-x_r y_j
)\frac{x_j}{h_j}\\
&=&\sum_{j\geq n} (L^{i-1}_1)_{n+1, j} \frac{x_j}{h_j}
-\sum_{r\geq j\geq n} (L^{i-1}_1)_{n+1, r} x_r
\left(\frac{1}{h_j}-\frac{1}{h_{j-1}}\right)\\
&=&\sum_{j\geq n} (L^{i-1}_1)_{n+1, j} \frac{x_j}{h_j}
-\sum_{r\geq n} (L^{i-1}_1)_{n+1, r} x_r
\left(\frac{1}{h_r}-\frac{1}{h_{n-1}}\right)\\
&=& \frac{1}{h_{n-1}} \sum_{r\geq n} (L^{i-1}_1)_{n+1,r} x_r
\end{eqnarray*}
Next establish (\ref{Ham-Lax2}), using
 \bea
% (L_2^{\al})_{ij} &\neq &0, \mbox{   when} ~~j\geq i-\al
%  \label{rel1}\\
 (L_2)_{n+1,n}&=&h_n/h_{n-1}=1-x_ny_n
  \label{rel2}\\
  (L_2)_{n+1,r} x_{j-1}
&=& x_n (L_2)_{jr}
 \mbox{ provided $r>n$ and $r>j-1$} \label{rel3}\eea
 Indeed, setting $i=k+1$, one computes
 \bea
 \lefteqn{
 \sum_{j=n-k}^{n} (L_2^{k+1})_{n+1,j}~ x_{j-1} -
  \frac{h_n}{h_{n-1}}
   \sum_{j=n-k}^{n+1} (L_2^{k})_{nj} ~x_{j-1}
    }  \no\\
  &=&
    \sum_{j=n-k}^{n}
    \sum_{r=n}^{j+k} (L_2)_{n+1,r} (L_2^{k})_{rj} x_{j-1}
 -
  \frac{h_n}{h_{n-1}}
   \sum_{j=n-k}^{n+1} (L_2^{k})_{nj} x_{j-1}
\no\\
&=& \frac{h_n}{h_{n-1}}\sum_{j=n-k}^{n}
    %(L_2)_{n+1,n}
    (L_2^{k})_{nj} x_{j-1}
 -
  \frac{h_n}{h_{n-1}}
   \sum_{j=n-k}^{n+1} (L_2^{k})_{nj} x_{j-1}
\no\\
&& \hspace{3cm}+
  \sum_{j=n-k}^{n}
    \sum_{r=n+1}^{j+k} (L_2)_{n+1,r} (L_2^{k})_{rj} x_{j-1}
 ~~~~~~~\mbox{using (\ref{rel2})}\no\\
 &\stackrel{*}{=}&
  x_n \left(-(L_2)_{n+1,n}
     (L_2^{k})_{n,n+1}
   +
    \sum_{j=n-k}^{n}~
    \sum_{r=n+1}^{j+k} (L_2)_{jr} (L_2^{k})_{rj}\right)\no\\
    &=&0,
 \label{previous}\eea
 using in $\stackrel{*}{=}$ formulas (\ref{rel2}), (\ref{rel3}) and the
 inequalities
    $n-k-1\leq j-1\leq n-1 <n+1 \leq r \leq j+k$.

As a last step, we show that indeed the last expression
(\ref{previous}) vanishes. The first expression in the bracket
equals
\bea
 \lefteqn{(L_2)_{n+1,n}
     (L_2^{k})_{n,n+1}}\no\\
     &=&
    (L_2)_{n+1,n}
     \sum_{
     {n \leq \beta_1+1\leq \beta_2+2 \leq \beta_3+3
     \leq }\atop{\ldots \leq \beta_{k-1}+k-1
     \leq n+k+1}}
     %\\&&\\&&\hspace{6cm}
     (L_2)_{n\beta_1}
              (L_2)_{\beta_1\beta_2}
              \ldots
              (L_2)_{\beta_{k-1},n+1}~,
  \no\\
  \label{sum-one}\eea
 whereas the second expression in the bracket of (\ref{previous})
 equals, upon setting $\al_0=r, \al_k=j$:
 \bea
\lefteqn{
    \sum_{j=n-k}^{n}~
    \sum_{r=n+1}^{j+k} (L_2)_{jr} (L_2^{k})_{rj}
    }
    \no\\
& =&
 \sum_{r=n+1}^{n+k}~
    \sum_{j=r-k}^{n} (L_2)_{jr} (L_2^{k})_{rj} \no\\
 &=&\sum_{
  {{n+1 \leq \al_0\leq \al_1+1\leq \al_2+2\leq }\atop
  {\ldots \leq
  \al_{k-1}+k-1\leq   \al_k+k\leq n+k}} %\atop {\al_k \leq \al_0+1}
  }
 % \\&&\\&&\hspace{2cm}
  (L_2)_{\al_0,\al_1}
  (L_2)_{\al_1,\al_2}\ldots
  (L_2)_{\al_{k-1},\al_k}
  (L_2)_{\al_k,\al_0}
  \no\\
  \label{sum-two}\eea
Each term in this sum is a product of $k+1$ entries of $L_2$, with
indices $\al_i+i$, squeezed between $n+1$ and $n+k$. Therefore, we
must have for some $ 1\leq j \leq k$ that
 $$ \al_{j}+j=  \al_{j+1}+j+1=n+j+1,$$
   and so for that $(\al_j,\al_{j+1})$, we have
   $$ (L_2)_{\al_j,\al_{j+1}}=(L_2)_{ n+1,n},$$
 which appears in every term of the sum (\ref{sum-two}); it can
 therefore be taken out,
  leaving sums of products of $k$ terms.
The inequalities under the summation sign of
 (\ref{sum-two}) can then be written as follows
{\footnotesize
$$  \hspace{1.5cm}{ {n+j+1}\atop {\|} }~~~
 {{n+j+1} \atop {\|} }$$
$$
 n+1\leq \al_0\leq \al_1+1\leq \ldots
  \leq \al_{j-1}+j-1 \leq \al_j+j=\al_{j+1}+j+1
  \leq \al_{j+2}+j+2 \leq \ldots \leq \al_k+k \leq n+k
 $$
}
 Add $k-j$ to the sequence above, from $n+1$ up to including $\al_j+j=n+j+1$ and
$-j-1$ to the sequence above, starting with $\al_{j+1}+j+1=
n+j+1$, up to $n+k$, yielding the two sequences
 $$n+k-j+1 \leq \al_0+k-j\leq \ldots \leq \al_{j-1}+k-1\leq n+k+1$$
and
 $$n \leq \al_{j+2}+1\leq \ldots \leq \al_k+k-j-1\leq n+k-j-1$$
Since obviously $n+k-j-1 < n+k-j+1$, we have the inequalities
appearing in the summation of formula (\ref{sum-one}), but with
$\beta$'s replaced by $\al$'s. This ends the proof of Proposition
1.1. \qed

According to \cite{AvM1}, we also have ($h:=\diag
(h_0,h_1,\ldots)=\diag
(\frac{\tau_1}{\tau_0},\frac{\tau_2}{\tau_1},\ldots) $)
 \be
\begin{array}{ll}
x_{n+1}y_{n+1}=1-\frac{h_{n+1}}{h_{n}}&~~
 ~y_{n+1} x_{n+1}=1- \frac{h_{n+1}}{h_n}\\
 \\
x_{n+1}y_n  =-\frac{\pl}{\pl t_1}\log
h_{n}&~~~y_{n+1}x_n =\frac{\pl}{\pl s_{1}}\log h_n\\
 \\
x_{n+1}y_{n-1}=-\frac{h_{n-1}}{h_n}
 \left(\frac{\pl}{\pl
t_1}\right)^2\log \tau_n&~~~y_{n+1}
x_{n-1}=-\frac{h_{n-1}}{h_n}\left(\frac{\pl}{\pl
s_1}\right)^2\log \tau_n
\end{array}
\label{tau-relations}
 \ee
  \bea
   x_{n+1}y_{n-k}&=&
 -\frac{h_{n-k}}{h_n}\frac{p_{k+1}(\tilde\pl_t)
\tau_{n-k+1}\circ\tau_n} {\tau_{n-k+1}\tau_n}
\nonumber\\ y_{n+1}x_{n-k}&=&
 -\frac{h_{n-k}}{h_n}\frac{p_{k+1}(-\tilde\pl_s)
\tau_{n-k+1}\circ\tau_n} {\tau_{n-k+1}\tau_n} ,~~k
\geq 0. \eea

\begin{lemma} The $t_i$- and $s_i$-derivatives
 of $x_n$ and $y_n$ can be expressed in terms
of the elements on the main diagonal and one above and
below the main diagonal of $L_k^i$ and $L_k^{i-1}$:
 \bean
  \frac{x_n y_n}{v_n}  \sum_{i\geq 1}
  \left( \al_i\frac{\pl}{\pl t_i}-
   \beta_i\frac{\pl}{\pl s_i}\right)
   \log x_n
 &=& y_n\frac{\pl }{ \pl y_n} \sum_{i\geq 1}
 \left( \al_iH_i^{(1)}-  \beta_i H_i^{(2)}  \right)\\
  &=&
   \sum_{i\geq 1}^{}
\left\{ \begin{array}{l}
  \al_i
  \Bigl(-( L_1^{i})_{n+1,n+1}+( L_1^{i-1})_{n+1,n}\Bigr)
    \\ \\
    +
  \beta_i
  \Bigl(  (L_2^i)_{nn} -  (L_2^{i-1})_{n,n+1}\Bigr)
  \end{array}\right\}
 \\ &&\\ &&\\
 \frac{x_n y_n}{v_n}  \sum_{i\geq 1}
  \left( \al_i\frac{\pl}{\pl t_i}-
   \beta_i\frac{\pl}{\pl s_i}\right)
   \log y_n
 &=&
 -x_n \frac{\pl }{ \pl x_n}\sum_{i\geq 1}
   \left( \al_iH_i^{(1)}-  \beta_i H_i^{(2)}  \right)
   \\
  &=&
   \sum_{i\geq 1}^{}
  \left\{
  \begin{array}{l}
  \al_i\left(( L_1^i)_{nn}-( L_1^{i-1})_{n+1,n}\right)
\\ \\-\beta_i
 \left((L_2^i)_{n+1,n+1}-(L_2^{i-1})_{n,n+1}\right)
 \end{array}
 \right\}  \\
  \eean \vspace{-1cm}\be\label{tderivative}\ee
In particular, we have
 \bea
  \frac{x_n y_n}{v_n}
  \left( c\frac{\pl}{\pl t_1}-
   a\frac{\pl}{\pl s_1}\right)
   \log x_n
 &=& y_n\frac{\pl }{ \pl y_n}(c
 H_1^{(1)}-aH_1^{(2)})    \nonumber \\
  &=& -ax_{n-1}y_{n}+cx_{n+1}y_n
\nonumber\\
  &=&   a(L_2)_{nn}-c(L_1)_{n+1,n+1}
 \nonumber\\
 \frac{x_n y_n}{v_n}
  \left( c\frac{\pl}{\pl t_1}-
   a\frac{\pl}{\pl s_1}\right)
   \log y_n
 &=& -x_n\frac{\pl }{ \pl x_n}(c
 H_1^{(1)}-aH_1^{(2)})\nonumber\\
  &=& ax_{n}y_{n+1}-cx_{n}y_{n-1} \nonumber\\
  &=&-a(L_2)_{n+1,n+1}+c(L_1)_{nn}
. \label{t1derivative}\eea
\end{lemma}

\proof The first equality in each of the identities
above follows immediately from the Hamiltonian vector
fields (\ref{Toeplitz}), with Hamiltonians
(\ref{Hamiltonians}). Note the following

\bigbreak

  \noindent
$\displaystyle{\left(x_n\frac{\pl}{\pl
x_n}-y_n\frac{\pl}{\pl y_n}\right)L_1}=$

\medbreak $$\hspace{63mm}\stackrel{n+1}{\downarrow}$$
$$\begin{array}{c} \\
\\
n\rg\\
 \\
\end{array}
\MAT{8}
 & & & & &0  \\
 &O& & & &\vdots& &O \\
 & & & & &0& \\
-x_ny_0&
&-x_ny_1&\ldots&-x_ny_{n-1}&-x_ny_n+x_ny_n&0&\ldots\\
 & & & & &x_{n+1}y_n\\
 & & & & &x_{n+2}y_n& &O\\
&O& & & &\vdots \mat $$

\bigbreak and  \bigbreak

\newpage

\noindent $\displaystyle{\left(x_n\frac{\pl}{\pl
x_n}-y_n\frac{\pl}{\pl y_n}\right)L_2}=$

\medbreak $$\hspace{1cm}\stackrel{n}{\downarrow}$$ $$
\begin{array}{c}
\\
%=
\\
\\
n+1\rg\\
 \\
\end{array}
\MAT{9}
 & & & & &x_0y_n& & & \\
 & &O& & &\vdots& &O& \\
 & & & & &x_{n-1}y_{n-1}& \\
0&\ldots&
&\ldots&0&x_ny_n-x_ny_n&-x_ny_{n+1}&-x_ny_{n+2}&\ldots\\
 & & & & &0\\
 & &O& & &\vdots& &O
\mat $$ \be\label{6} \ee
We shall also need the following trivial identities:
$$\left( L_1^i\right)_{nn}=\left\{
\begin{array}{l}
 \displaystyle{ \left( L_1
  L_1^{i-1} \right)_{nn}=
  -x_n\sum_{j=1}^{n+1}y_{j-1}
   \left(  L_1^{i-1}\right)_{jn}
    +\left(  L_1^{i-1}\right)_{n+1,n}
     } \\  \\
%
%    $$
   \displaystyle{
   \left(  L_1^{i-1}L_1  \right)_{nn}=
  -y_{n-1}\sum_{j=n-2}^{n+i-2}
  x_{j+1} \left(  L_1^{i-1}  \right)_{n,j+1}
  +\left(  L_1^{i-1}\right)_{n,n-1}
  }
  \end{array} \right.
  $$
and

  \be \left( L_2^i\right)_{nn}=\left\{ \begin{array}{l}
 \displaystyle{ \left( L_2
  L_2^{i-1} \right)_{nn}=
  -x_{n-1}\sum_{j=n-2}^{n+i-2}y_{j+1}
   \left( L_2^{i-1}\right)_{j+1,n}
    +\left( L_2^{i-1}\right)_{n-1,n}
     } \\  \\
%
%    $$
   \displaystyle{
   \left(  L_2^{i-1}L_2  \right)_{nn}=
  -y_{n}\sum_{j=1}^{n+1}
  x_{j-1} \left(  L_2^{i-1}  \right)_{n,j}
  +\left( L_2^{i-1}\right)_{n,n+1}
  }
  \end{array} \right.
 \label{7} \ee
We now have (see the definition (\ref{Hamiltonians})
of
 $H_i^{(k)}$)
 \bean
&& y_n\frac{\pl }{ \pl y_n}
  \sum_{i\geq 1}\left( \al_iH_i^{(1)}-
  \beta_i H_i^{(2)}  \right)  \\
   &=&
  -y_n\frac{\pl }{ \pl y_n}
  \sum_{i\geq 1}\tr \left( \frac{\al_i}{i}
  L_1^{i}- \frac{\beta_i}{i}
  L_2^{i}
  \right)
 \\&&\\
 &=&
 \left\la
\sum_{i\geq 1}{}\al_i  L_1^{i-1},
 -y_n\frac{\pl }{ \pl
 y_n}  L_1
  \right\ra
  -
  \left\la
\sum_{i \geq 1}{}\beta_iL_2^{i-1},
 -y_n\frac{\pl }{ \pl
 y_n}L_2
  \right\ra
 \\&&\\&&
  \\
 &=&
  \sum_{i\geq 1}^{}
  \al_i
  \left(
  y_n \sum_{j=n-1}^{n+i-1}x_{j+1}
   (  L_1^{i-1})_{n+1,j+1}
  \right)
  +
   \sum_{i\geq 1}^{}
  \beta_i
  \left(
 -y_n \sum_{j=1}^{n+1}x_{j-1} (L_2^{i-1})_{n,j}
   \right)
  \\&&\hspace{9cm}\mbox{using (\ref{6})}\\&&
 \\&&\\&=&
\sum_{i\geq 1}^{}
  \al_i
  \left(-(  L_1^{i})_{n+1,n+1}+(  L_1^{i-1})_{n+1,n}\right)
+\sum_{i\geq 1}^{}
  \beta_i
  \left(  (L_2^i)_{nn} -  (L_2^{i-1})_{n,n+1}\right),
 \\
  \eean
  using in the last equality identities (\ref{7}).

Similarly, one computes
  \bean
 \lefteqn{-x_n \frac{\pl }{ \pl
  x_n}
  \sum_{i\geq 1}\left( \al_iH_i^{(1)}-
  \beta_i H_i^{(2)}  \right)}\\
   &=&
  x_n  \frac{\pl }{ \pl
  x_n}
  \sum_{i\geq 1}\tr \left( \frac{\al_i}{i}
    L_1^{i}- \frac{\beta_i}{i}
  L_2^{i}
  \right)
 \\&&\\
 &=&
 \left\la
\sum_{i\geq 1}{}\al_i  L_1^{i-1},
 x_n  \frac{\pl }{ \pl x_n}  L_1
  \right\ra
  -
  \left\la
\sum_{i \geq 1}{}\beta_iL_2^{i-1},
 x_n  \frac{\pl }{ \pl x_n}L_2
  \right\ra
 \\&&\\
 &=&
  \sum_{i\geq 1}^{}
  \al_i
  \left( -x_n \sum_{j=1}^{n+1}y_{j-1}
  (  L_1^{i-1})_{j,n}
  \right)
  +
   \sum_{i\geq 1}^{}
  \beta_i
  \left(
  x_n \sum_{j=n-1}^{n+i-1}y_{j+1} (L_2^{i-1})_{j+1,n+1}
  \right)
  \\&&\\&=&
  \sum_{i\geq 1}^{}
  \al_i\left(( L_1^i)_{nn}-( L_1^{i-1})_{n+1,n}\right)
+\sum_{i\geq 1}^{}\beta_i
 \left(-(L_2^i)_{n+1,n+1}+(L_2^{i-1})_{n,n+1}\right)
 \\
  \eean

  The last couple of relations (\ref{t1derivative})
   follow from specializing (\ref{tderivative}) to
   $i=1$,  thus ending the proof of Lemma 1.2.\qed

   \begin{lemma}
    \bean
\left(\frac{\pl  L^i_1 }{\pl t_1}\right)_{nn}&=&v_n(
L^i_1)_{n+1,n}-v_{n-1} ( L^i_1)_{n,n-1}\\
 %\left(\frac{\pl\hat L_1 }{\pl
%t_1}\right)_{n,n}&=&-v_n x_{n+1}y_{n-1}+
%v_{n-1}x_ny_{n-2},\\
  \left(\frac{\pl  L_2^i }{\pl
t_1}\right)_{nn}&=&(L^i_2)_{n+1,n}-(L_2^i)_{n,n-1}
 \\
 \left(\frac{\pl  L^i_1 }{\pl
s_1}\right)_{nn}&=&(  L^i_1)_{n-1,n}- (
L^i_1)_{n,n+1}%~~\mbox{CHECK!}
 \\
 %\left(\frac{\pl\hat L_1 }{\pl
%t_1}\right)_{n,n}&=&-v_n x_{n+1}y_{n-1}+
%v_{n-1}x_ny_{n-2},\\
  \left(\frac{\pl  L_2^i }{\pl
s_1}\right)_{nn}&=&v_{n-1}(L^i_2)_{n-1,n}
-v_n(L_2^i)_{n,n+1} \eean
   \end{lemma}

   \proof From the Toda equations (\ref{Toda}) for
   $\hat L_i$, one computes the $t_1$-flow for $\hat L_1$,
   where $h:=\diag (h_1,h_2,\ldots)$ and where
   $A_+$, $A_{++}$ and $A_0$ denote the
   upper-triangular, strictly upper-triangular and
   diagonal part of the matrix $A$, (remember
   $L_{1}=h^{-1}\hat L_1 h $)
   \bean
    \frac{\pl L_1^i}{\pl t_1}
     &=&\frac{\pl}{\pl t_1}(h^{-1}\hat L_1^i h)
     \\
 &=&
     h^{-1}\frac{\pl \hat L_1^i}{\pl t_1}h
     -\frac{\pl\log h }{\pl t_1}h^{-1}
      \hat L_1^i h+h^{-1}\hat L_1
     ^i h
      \frac{\pl\log h }{\pl t_1}
      \\
 &=&
  [h^{-1}(\hat L_1)_+h,h^{-1}\hat L_1^i h]-
   \left[\frac{\pl\log h}{\pl t_1},h^{-1}\hat L_1^i h\right]
 \\
  &=&
   \left[(
L_1)_+-\frac{\pl\log h }{\pl t_1}, L_1^i\right]
 \\
 &=&
  [( L_1)_{++}, L_1^i],~~~
   \mbox{using (\ref{tau-relations})}\\
  &=&
  [\diag (v_1,v_2,\ldots) \Lb, L_1^i].
 \eean
 Hence, setting $v_n=h_n/h_{n-1}=1-x_ny_n$
 $$ \left(\frac{\pl L^i_1 }{\pl
t_1}\right)_0=\diag\left(v_1( L^i_1)_{21}, v_2(
L^i_1)_{32}-v_1( L^i_1)_{21},\ldots\right).
  $$
 In particular,
 \bean
\left(\frac{\pl  L^i_1 }{\pl t_1}\right)_{nn}&=&v_n(
L^i_1)_{n+1,n}-v_{n-1} (  L^i_1)_{n,n-1}\\
\left(\frac{\pl  L_1 }{\pl t_1}\right)_{n,n}&=&-v_n
x_{n+1}y_{n-1}+ v_{n-1}x_ny_{n-2}.
  \eean
    We also need the $t_1$-derivative of $L_2^i$,
  \bean \frac{\pl L_2^i
}{\pl t_1}&=&[(\hat L_1)_+,L_2^i]\\
&=&\left[\left(\begin{array}{cccccc} -x_1y_0& &1& \\
 & & & & &O\\
 & &-x_2y_1&1\\
 & & &-x_3y_2&1\\
 &O& & &\ddots&\ddots
\end{array}
\right),L_2^i\right] \eean
 hence
  $$ \left(\frac{\pl
L_2^i }{\pl
t_1}\right)_0=\diag\left((L^i_2)_{21},(L_2^i)_{3,2}-(L_2^i)_{2,1},(L^i_2)_{4,3}-
(L_2^i)_{3,2}, \ldots\right) ,$$
  leading to
\bean \left(\frac{\pl  L_2^i }{\pl
t_1}\right)_{n,n}&=&(L^i_2)_{n+1,n}-(L_2^i)_{n,n-1}\\
\left(\frac{\pl  L_2 }{\pl
t_1}\right)_{n,n}&=&x_{n-1}y_{n-1}-x_ny_n ,\eean
while the latter relations of Lemma 1.3 are obtained from the
first two by the 2-Toda involution (\ref{invol}). This ends the
proof of Lemma 1.3.\qed

%\newpage

\subsection{Virasoro constraints}

According to \cite{AvM1}, the (vector) vertex
operator\footnote{For
 $v=(v_{0},v_{1},\ldots)^{\top},~
  (\Lambda v)_{n}=v_{n+1},
  ~(\Lambda^{\top}v)_{n}=v_{n-1}$, and $\chi(z):=(1,z,z^2,...)$.}
\be
\BX_{12}(t,s;u,v)=\Lb^{-1}e^{\sum_1^{\iy}(t_i u^i-s_i
v^{i})} e^{-\sum_1^{\iy}(\frac{ u^{-i}}{i}
\frac{\pl}{\pl t_i}-
 \frac{ v^{-i}}{i}
\frac{\pl}{\pl s_i})}\chi(uv), \label{vertex}\ee
 acting on vectors of functions of $t$ and $s$,
  interacts with the operators
 $\BJ_k^{(i)}(t)=
 \left(\BJ_{k,n}^{(i)}(t,n)\right)_{n\geq 0}$,
 as follows: (for definitions, see the appendix 1)
  \bea
  u^{k}\BX_{12}(t,s;u,v)&=& \left[
\BJ_k^{(1)}(t), \BX_{12}(t,s;u,v)\right] \nonumber\\
\frac{\pl}{\pl u}u^{k+1}\BX_{12}(t,s;u,v)&=& \left[
\BJ_k^{(2)}(t), \BX_{12}(t,s;u,v)\right].
 \label{diffvertex}\eea
 A similar statement can
be made, upon replacing the operators $u^k$ and
$\frac{\pl}{\pl u}u^{k+1}$ by $v^k$ and
$\frac{\pl}{\pl v}v^{k+1}$, and upon using
$\tilde\BJ_k^{(i)}(s)=\BJ_k^{(i)}(-s)$.

 Also
consider the vertex operator, integrated over the unit
circle and depending on
an integer $\gamma$, %and a weight $\rho$ on
 \bea
  {\BY}^{\gamma}_{}(t,s)&=&\int_{S^1} \frac{du}{2\pi i u}
   u^{\gamma}\BX_{12}(t,s;u,u^{-1}).
    \label{intvertex}\eea
and the vector Virasoro constraint $
{\BV}^{\gamma}_{k}(t,s):= (
{\BV}^{\gamma}_{k,n})_{n\geq 0}$
 \be
  \BV^{\gamma}_k:=\BJ_{k}^{(2)}(t)-
  \BJ_{-k}^{(2)}(-s)
-(k-\gamma)\bigl( \theta \BJ_{k}^{(1)}(t)+(1-\theta)
 \BJ_{-k}^{(1)}(-s)
 \bigr)
 \ee
depending on a free parameter $\theta$. \vspace*{.5cm}

    \begin{theorem} {\em (Adler-van Moerbeke \cite{AvM1})}
The multiple integrals %$I=(I_0=1,I_1,...)$,
 over the unit circle $S^1$,
\be
\tau^{\gamma}_n(t,s)=
\frac{1}{n!}\int_{(S^1)^{n}}|\Dt_n(z)|^{2}
 \prod_{k=1}^n
z_k^{\gamma}e^{\sum_1^{\iy}(t_i z_k^i-s_iz_k^{-i})}
 \frac{dz_k}{2\pi i z_k},~~~n>0,\label{tau}\ee with $I_0=1$,
 satisfy an SL(2,$\BZ$)-algebra of Virasoro constraints:
\be
\BV^{\gamma}_{k,n}\tau_n(t,s)=0,
 ~~~\mbox{for}~ \left\{\begin{array}{l} k
=-1 ,~\theta= 0\\ k=0 , ~~\theta~~ {arbitrary}\\
  k=1 ,~
\theta= 1\end{array}\right\} \mbox{\,\,only.
 }
 \ee

\end{theorem}

Working out the Virasoro equations of Appendix 1 (for
$\beta=1/2$)
  \bea
 \lefteqn{ {\BV}^{\gamma}_{k,n}}\no\\
  &:=&\BJ_{k,n}^{(2)}(t,n)-
  \BJ_{-k,n}^{(2)}(-s,n) -(k-\gamma)\left( \theta
\BJ_{k,n}^{(1)}(t,n)+(1-\theta)\BJ_{-k,n}^{(1)}(-s,n)
   \right) \no\\
 &=&\frac{1}{2}\left(J^{(2)}_{k}(t)-J^{(2)}_{-k}(-s)+
 (2n+k+1)J^{(1)}_{k}(t)-(2n-k+1)J^{(1)}_{-k}(-s)\right)
 \no\\ &
&\hspace{1cm}-(k-\gamma)\left(\theta
J^{(1)}_{k}(t)+(1-\theta)J^{(1)}_{-k}(-s)\right)+\gamma
n\dt_{k,0},\label{Vir}
 \eea
 one finds
 Virasoro constraints for the integral
 $\tau_n^{\gamma}$:
{\footnotesize
\begin{eqnarray}
{\BV}^{\gamma}_{-1,n}\tau_n^{\gamma}&=&\left(\sum_{i\geq
1}(i+1)t_{i+1}\frac{\pl}{\pl t_{i}}-\sum_{i\geq
2}(i-1)s_{i-1}\frac{\pl}{\pl
s_{i}}+nt_1+(n-\gamma)\frac{\pl}{\pl
s_{1}}\right)\tau_n^{\gamma}=0\nonumber\\
{\BV}^{\gamma}_{0,n}\tau_n^{\gamma}&=&\sum_{i\geq
1}\left(it_{i}\frac{\pl}{\pl
t_{i}}-is_{i}\frac{\pl}{\pl s_{i}}\right)\tau_n^{\gamma}+\gamma
n \tau_n^{\gamma} =0
\label{Virasoro}\\
{\BV}^{\gamma}_{1,n}\tau_n^{\gamma}&=&\left(-\sum_{i\geq
1}(i+1)s_{i+1}\frac{\pl}{\pl s_{i}}+\sum_{i\geq
2}(i-1)t_{i-1}\frac{\pl}{\pl
t_{i}}+ns_1+(n+\gamma)\frac{\pl}{\pl t_1}
\right)\tau_n^{\gamma}=0.\nonumber\\\no
\end{eqnarray}
}%\be \vspace{-1.5cm} \label{Virasoro}\ee

The theorem is based on two lemmas:

 \begin{lemma} The following commutation relations
 hold:
 \be
 %\frac{1}{u^{\pm 1}}
u^{-\gamma} u\frac{d}{du}u^{k+\gamma}
\BX_{12}(t,s;u,u^{-1}) =\left[ {\BV}_k^{\gamma}
  ,\BX_{12}(t,s;u,u^{-1})\right],
   \label{lemma1}
   \ee and
\be
   \left[{\BY^{\gamma}}, \BV^{\gamma}_k
 \right]=0.\label{commutation}
 \ee

 \end{lemma}

\proof Using (\ref{diffvertex}), a standard
computation shows
\begin{eqnarray}
\lefteqn{
 u\frac{d}{du}u^k \BX_{12}(t,s;u,u^{-1})}
  \\ &=&\left( u^{k+1} \frac{d}{du}+ku^{k}\right)
 \BX_{12}(t,s;u,u^{-1})\nonumber\\
&=&\left. \left(u^{k+1}\frac{\pl}{\pl u}-v^{1-k}
 \frac{\pl}{\pl v}+ku^k\right) \BX_{12}
 (t,s;u,v) \right|_{v=u^{-1}}\nonumber\\
  &=&\left. \left(\frac{\pl}{\pl u}u^{k+1}-\frac{\pl}{\pl v}v^{1-k}
 -ku^k\right)\BX_{12}(t,s;u,v) \right|_{v=u^{-1}}\nonumber\\
 &=&\left. \left(\frac{\pl}{\pl u}u^{k+1}-\frac{\pl}{\pl v}v^{1-k}
 -k\theta u^k-k(1-\theta)v^{-k}\right)\BX_{12}(t,s;u,v)
   \right|_{v=u^{-1}}\nonumber\\
  &=&\Bigl[ \BJ_{k}^{(2)}(t)-\BJ_{-k}^{(2)}(-s)
-k\left( \theta \BJ_{k}^{(1)}(t)+(1-\theta)~~
 \BJ_{-k}^{(1)}(-s) \right),\nonumber\\
   & &\hspace{5cm}\BX_{12}(t,s;u,u^{-1})\Bigr]
    \nonumber\\
&=&\left[ {\BV}_k^{(0)},\BX_{12}(t,s;u,u^{-1})\right],
   \end{eqnarray}
from which (\ref{lemma1}) follows, for $\gamma=0$.
More generally, we compute
\begin{eqnarray}
\lefteqn{ %\frac{1}{u^{\pm 1}}
u^{-\gamma} u\frac{d}{du}u^{k+\gamma}
\BX_{12}(t,s;u,u^{-1})}\\ &=&\left( u^{k+1}
\frac{d}{du}+ku^{k}\right)
 \BX_{12}(t,s;u,u^{-1})+\gamma  u^k\BX_{12}(t,s;u,u^{-1})\nonumber\\
&=&\left[ {\BV}_k^{(0)},\BX_{12}(t,s;u,u^{-1})\right] +\gamma
\left.\left( \theta u^k+(1-\theta)~~
 v^{-k} \right)\right|_{v=u^{-1}}\BX_{12}\nonumber\\
 &=&\left[ {\BV}_k^{(0)}
 +\gamma \left( \theta\BJ_{k}^{(1)}(t)
  +(1-\theta)\BJ_{-k}^{(1)}(-s) \right)
  ,\BX_{12}(t,s;u,u^{-1})\right]  \nonumber\\
&=&\left[ {\BV}_k^{\gamma}
  ,\BX_{12}(t,s;u,u^{-1})\right]
   \end{eqnarray}
from which (\ref{lemma1}) follows.
 We then have
  \bean \left[ \BV^{\gamma}_k,  {\BY}^{\gamma}(t,s)
 \right]&=&
\left[ \BV^{\gamma}_k, \int_{S^1}
\BX_{12}(t,s;u,u^{-1})
 u^{\gamma}\frac{du}{2\pi i u} \right]\\&=&
  \int_{S^1} \left[\BV_k^{\gamma}, \BX_{12}(t,s;u,u^{-1})
\right] u^{\gamma} \frac{du}{2\pi i u}  \\&=&
   \int_{S^1}\frac{du}{2\pi i} ~\frac{d}{d u}u^{k+\gamma
   }
  {\BX_{12}(t,s;u,u^{-1})}  \\ & =&0,
\eean leading to (\ref{commutation}). \qed

\begin{lemma} The vector $I:=(I_n)_{n\geq 0}$, with
$I_n=n! \tau_n^{\ga}$, is a fixed point for the vertex
operator $\BY^{\gamma}$,
\be
 {\BY^{\gamma}}_{}(t,s)I(t,s)=I(t,s).
 \label{fixedpoint}\ee

\end{lemma}

\proof Setting $\rho(dz)=z^{\gamma}dz$, one computes ,
for $n\geq 1$,
 \bea I_n(t,s)&=&n!\tau_n^{\ga}(t,s)\nonumber\\
&=&\int_{(S^1)^{n}}|\Dt_n(z)|^{2}
 \prod_{k=1}^n
\left(e^{\sum_1^{\iy}(t_i z_k^i-s_iz_k^{-i})}
 \frac{\rho(dz_k)}{2\pi i z_k}\right)
\nonumber\\ &&\nonumber\\
 &=&\int_{S^1}\frac{\rho(du)}{2\pi iu}e^{\sum_1^{\iy}(t_i u^i-s_i u^{-i})}
 u^{n-1}u^{-n+1}\nonumber\\
 &&\int_{(S^1)^{n-1}} \Dt_{n-1}(z)\bar \Dt_{n-1}(z)
 \no\\
 &&\hspace{1cm} \prod_{k=1}^{n-1}\left(1-\frac{z_k}{u}\right)
  \left(1-\frac{u}{z_k}\right)
e^{\sum_1^{\iy}(t_i z_k^i-s_i z_k^{-i})}
\frac{\rho(dz_k)}{2\pi i z_k}\nonumber\\
&=&\int_{S^1}\frac{\rho(du)}{2\pi
iu}e^{\sum_1^{\iy}(t_i u^i-s_i u^{-i})} ~e^{-
\sum_1^{\iy}\left(\frac{u^{-i}}{i}
 \frac{\pl}{\pl t_i}
 -\frac{u^{i}}{i}\frac{\pl}{\pl s_i}\right)}
 \nonumber\\
&&~~~~~\int_{(S^1)^{n-1}}
  \Dt_{n-1}(z)\bar \Dt_{n-1}(z)
 \prod_{k=1}^{n-1}
e^{\sum_1^{\iy}(t_i z_k^i-s_i z_k^{-i})}
\frac{\rho(dz_k)}{2\pi iz_k}\nonumber\\
&=&\int_{S^1}\frac{\rho(du)}{2\pi
iu}e^{\sum_1^{\iy}(t_i u^i-s_i u^{-i})} e^{-
\sum_1^{\iy}\left(\frac{u^{-i}}{i}\frac{\pl}{\pl t_i}
 -\frac{u^{i}}{i}\frac{\pl}{\pl s_i}\right)}
 I_{n-1}(t,s)
%&=&\int_{\BR}du\rho(u)I_E(u)|u|^{\beta (n-1)}e^{\sum_1^{\iy}t_i
%u^i}
%  e^{-\beta \sum_1^{\iy}\frac{u^{-i}}{i}
%\frac{\pl}{\pl t_i}}\tau_{n-1}(t)
\nonumber\\ &&\nonumber\\
 &=&\Big({\BY}^{\gamma}_{}(t,s)I(t,s)\Big)_n,
  \eea
   from which (\ref{fixedpoint}) follows. \qed

 {\medskip\noindent{\it Proof of Theorem 1.4:\/} }
 From (\ref{commutation}), we have
   \bean 0&=&\left([{\BV}^{\gamma}_{k},({\BY^{\gamma}})^n]
     I\right)_n\\ &=&\left({\BV}^{\gamma}_{k}
       (\BY^{\gamma})^n I-(\BY^{\gamma})^n
        {\BV}^{\gamma}_{k}I\right)_n\\
&=&\left({\BV}^{\gamma}_{k}I-(\BY^{\gamma})^n
  {\BV}^{\gamma}_{k}I\right)_n.
\eean
 Taking the $n^{{\rm th}}$ component and taking
into account the presence of $\Lambda^{-1}$ in
$\BX_{12}(t,s;u,u^{-1})$, we find
 \bean
0&=&\left({\BV}^{\gamma}_{k}I-{\BY}^n
 {\BV}^{\gamma}_{k}I\right)_{n}\\
 &=&{\BV}^{\gamma}_{k}I_{n}-\int_{S_{1}}\frac{du}{2\pi
iu}e^{\sum_{1}^{\iy}(t_{i}u^i-s_{i}u^{-i})}
e^{-\sum_{1}^{\iy}\left(\frac{u^{-i}}{i}
\frac{\pl}{\pl t_{i}}- \frac{u^{i}}{i} \frac{\pl}{\pl
s_{i}}\right)}
\\
&&   \hspace{2cm} \ldots \int_{S_{1}}\frac{du}{2\pi
iu} e^{\sum_{1}^{\iy}(t_{i}u^i-s_{i}u^{-i})}
e^{-\sum_{1}^{\iy}\left(\frac{u^{-i}}{i}
\frac{\pl}{\pl t_{i}}- \frac{u^{i}}{i} \frac{\pl}{\pl
s_{i}}\right)} {\BV}^{\gamma}_{k}I_{0}. \eean
  Remember from (\ref{Vir}),
${\BV}_{k}^{\gamma}(t,s)$ has the following form
 \bean
{\BV}^{\gamma}_{k}(t,s)&=&
\frac{1}{2}\left(J^{(2)}_{k}(t)-J^{(2)}_{-k}(-s)+
 (2n+k+1)J^{(1)}_{k}(t)-(2n-k+1)
 J^{(1)}_{-k}(-s)\right)\\
 & &\hspace{1cm}-(k-\gamma)\left(\theta
J^{(1)}_{k}(t)+(1-\theta)J^{(1)}_{-k}(-s)\right)+\gamma
n\dt_{k,0} . \eean
    and one checks immediately that,
given $\tau_0=1$,
 $$
{\BV}^{\gamma}_{k}(t,s)\tau_0=0\quad\mbox{only for}~~
\left\{\begin{array}{l} k =-1 ,~~ \theta= 0\\ k=0 ,
~~\theta~~ {arbitrary}\\
  k=1 ,~~
\theta= 1\end{array}\right\} , $$ ending the proof of Theorem
1.4.\qed

%\newpage

\section{Rational recursion relations}
\subsection{Weights}

\begin{lemma}
\bea \LR&=&\left\{
\begin{array}{l}
it_i=it_i^{(0)}:=\left\{\begin{array}{l}
u_i-(\gamma'_1d_1^i+\gamma'_2d_2^i),\mbox{~for~}1\leq
i\leq N_1\\
 \\
-(\gamma'_1d_1^i+\gamma'_2d_2^i),\mbox{~for~}N_1+1\leq
i<\infty
\end{array}\right.\\
 \\
is_i=is_i^{(0)}:=\left\{\begin{array}{l} -u_{-i}+
(\gamma''_1d_1^{-i}+\gamma''_2d_2^{-i}),\mbox{~for~}1\leq
i\leq N_2\\
 \\
(\gamma''_1d_1^{-i}+\gamma''_2d_2^{-i}),\mbox{~for~}N_2+1\leq
i<\infty
\end{array}\right.
\end{array}\right\}\nonumber\\ \label{locus}
\eea
 Then, setting $\gamma_1=\gamma'_1+\gamma''_1$ and
$\gamma_2=\gamma'_2+\gamma''_2$, we have
\bea
\lefteqn{e^{\sum^{\infty}_1(t_iz^i-s_iz^{-i})}\Bigl|_{\LR}
 }\nonumber\\&=&
 e^{P_1(z)+P_2(z^{-1})}%e^{\sum^{N_1}_{-N_2}u_iz^i/i}
  (1-d_1z)^{\gamma'_1}
   (1-d_2z)^{\gamma'_2}(1-d_1^{-1}z^{-1})^{\gamma''_1}
   (1-d_2^{-1}z^{-1})^{\gamma''_2}\nonumber\\&&
 \nonumber\\
&=&kz^{-\gamma''_1-\gamma''_2}
 e^{P_1(z)+P_2(z^{-1})}
(1-d_1z)^{\gamma_1}(1-d_2z)^{\gamma_2}
 %e^{\sum^{N_1}_{-N_2}u_iz^i},
  \label{2.1.2}\eea
 with a constant $k$ and
 \be
  P_1(z):=\sum_1^{N_1} \frac{u_iz^i}{i}
  ~~\mbox{and}
  ~~
  P_2(z^{-1}):=\sum_1^{N_2} \frac{u_{-i}z^{-i}}{i}
  .\ee
 Moreover, there exist $a,b,c$ such that
\bea
 a(i+1)t^{(0)}_{i+1}+bit_i^{(0)}
 +c(i-1)t_{i-1}^{(0)}
 &=&0~~~\mbox{for all~}i\geq N_1+2\no\\
 a(i-1)s^{(0)}_{i-1}+bis_i^{(0)}
 +c(i+1)s_{i+1}^{(0)}&=&0~~~\mbox{for all~}i\geq N_2+2
 .\label{linearsystem}\eea
 Then
 \bea
%\lefteqn{\hspace{-2cm}
 a(i+1)t^{(0)}_{i+1}+bit_i^{(0)}
 +c(i-1)t_{i-1}^{(0)}
&=& au_{i+1} +bu_i + cu_{i-1}
 +c\dt_{i1}(\gamma^{\prime}_1+ \gamma^{\prime}_2),\no\\
 && \hspace{2cm}~~\mbox{for~}1\leq i\leq N_1+1
 \no\\ \no\\
%
%\lefteqn{\hspace{-2cm}
 a(i-1)s^{(0)}_{i-1}+bis_i^{(0)}
 +c(i+1)s_{i+1}^{(0)}
  &=& -cu_{-i-1} -bu_{-i} -a u_{-i+1}
   -a\dt_{i1}(\gamma^{\prime\prime}_1+
    \gamma^{\prime\prime}_2)
   \,\no\\
 && \hspace{2cm}~~\mbox{for~}1\leq i\leq N_2+1
.\no\\ \label{relations}
 \eea
upon setting
$u_0=u_{N_1+1}=u_{N_1+2}=u_{-N_2-1}=u_{-N_2-2}=0$.

\begin{itemize}
  \item {\bf Case 1}. $d_1,d_2,d_1-d_2\neq 0$ and
  $|\gamma_1^{\prime}|
   + |\gamma_1^{\prime\prime}| ,
   ~|\gamma_2^{\prime}| +|\gamma_2^{\prime\prime}|
   \neq 0$. Then the unique
  solution to (\ref{linearsystem}) is given by $$
  a=1,~ b=-d_1-d_2,~ c=d_1d_2.$$

  \item {\bf Case 2}. $d_1\neq 0,~\gamma_1^{\prime}\neq 0
  $ arbitrary, $d_2= 0,
  \gamma^{\prime}_2=
  \gamma^{\prime\prime}_1=
  \gamma^{\prime\prime}_2=0.
  $
  Then there exist two solutions
  $$(a,b,c)= (1,-d_1,0)~~ \mbox{and}~~
   (a,b,c)= (0,1,-d_1)$$
  such that (\ref{linearsystem}) holds.
  \item  {\bf Case 3}. $d_1=d_2=0, ~
   \gamma_1^{\prime}=
   \gamma_2^{\prime}=
   \gamma_1^{\prime\prime}=
   \gamma_2^{\prime\prime}=0 $. Then $a,b,c$ may be
   taken arbitrary.

\end{itemize}

%\vspace{2cm}

\end{lemma}

\proof Formula (\ref{2.1.2}) follows immediately from
 $1-x=\mbox{exp}~(-\sum_1^{\iy}x^i/i)$, while
 (\ref{linearsystem}) and (\ref{relations}) are
 obvious. \qed

\remark The locus $\LR$, defined in (\ref{locus}), provides
 the only example where (\ref{linearsystem}) holds

\subsection{Rational recursion relations}

Considering the $t$-dependent basic variables,
\be x_n(t)=(-1)^n \frac{\tau_n^{+}(t)}{\tau_n (t)}
~~~~ \mbox{and} ~~~~ y_n(t)= (-1)^n
\frac{\tau_n^{-}(t)}{\tau_n(t)}
 ,\label{tbasicvariables}\ee
where $\tau_n$ is the integral
\bean \tau_n^{\pm}&=&  \frac{1}{n!}
\int_{(S^1)^{n}}|\Dt_n(z)|^{2}
 \prod_{k=1}^n
\left(z_k^{\gamma\pm 1}e^{\sum_1^{\iy}(t_i
z_k^i-s_iz_k^{-i})}
 \frac{dz_k}{2\pi i z_k}\right)
\\
  \tau_n^{}&=&  \frac{1}{n!}
\int_{(S^1)^{n}}|\Dt_n(z)|^{2}
 \prod_{k=1}^n
\left(z_k^{\gamma}
 e^{\sum_1^{\iy}(t_i
z_k^i-s_iz_k^{-i})}
 \frac{dz_k}{2\pi i z_k}\right).
 \eean
Along the locus $\LR$, defined in (\ref{locus}), these
integrals and variables reduce to the original
integrals and variables (\ref{integral}) and
(\ref{basicvariables}). In the statement below, we
deal with the variables $x_n(t)$ and $y_n(t)$,
without restricting to the locus $\LR$.

\begin{theorem}
Set $v_n:=1-x_ny_n$,
  \bean
\alpha_i(t)&:=&a(i+1)t_{i+1}+bit_i+c(i-1)t_{i-1}
 +c(n+\gamma)\dt_{i1}
  \\
\beta_i(s)&:=&a(i-1)s_{i-1}+bis_i+
 c(i+1)s_{i+1}
 -a(n-\gamma)\dt_{i1}
,\eean
  and
  \be
  \LR_1^{(n)}=\sum_{i\geq 1} \al_i(t)  L_1^i
   ~~~\mbox{and}~~\LR_2^{(n)} =-\sum_{i\geq 1}\beta_i (t)
   L_2^i.
   \label{LR}\ee
 Then the following holds:

 \medbreak

\noindent $\bullet$ CASES 1 and 2: $a,c$ not both $=0$:
  \be
 \pl_n (\LR_1^{(n)}
   -\LR_2^{(n)})_{n,n}
  ~~~~~+~~~~~(c L_1-aL_2)_{nn}=0
  \label{identity1}\ee
and
 \be
 \pl_n^2\left(v_{n-1}\LR_1^{(n-1)}-\LR_2^{(n-1)}
 \right)_{n,n-1}
  +\pl_n \left(c( L_1^2)_{nn}+b( L_1)_{nn}\right)
  =0.
    \label{identity2}
  \ee

\medbreak

 \noindent $\bullet$ CASE 3: $a=c=0,~b=1$:
\be
\left\{ \begin{array}{l}
 \displaystyle{ \sum_{i\geq 1}^{}\left\{\al_i
  \left(( L_1^{i})_{n+1,n+1}-
   ( L_1^{i-1})_{n+1,n}\right)
    -
  \beta_i
  \left(  (L_2^i)_{nn} -  (L_2^{i-1})_{n,n+1}\right)
  \right\}}\\ \hspace{9cm}-n\frac{x_ny_n}{v_n} =0 \\  \\
 \displaystyle{ \sum_{i\geq 1}^{} \left\{
  \al_i\left(( L_1^i)_{nn}-( L_1^{i-1})_{n+1,n}
  \right)   -\beta_i
 \left((L_2^i)_{n+1,n+1}-(L_2^{i-1})_{n,n+1}\right)
 \right\}}\\ \hspace{9cm}-n\frac{x_ny_n}{v_n}=0
  \end{array}\right.  \label{identity3}
  \ee

\end{theorem}

\remark Written out, the equations (\ref{identity1})
and (\ref{identity2}) take on the following form
 \be
   (\LR_1-\LR_2+aL_2-cL_1)_{nn}-
  (\LR_1-\LR_2-aL_2+cL_1)_{n+1,n+1}
  =0 \label{identity1'}
   \ee
   and
   \bea
  \lefteqn{  2(v_n\LR_1-\LR_2)_{n+1,n}
  -(v_{n-1}\LR_1-\LR_2)_{n,n-1}
  -(v_{n+1}\LR_1-\LR_2)_{n+2,n+1}}\no\\
  && +a(v_{n+1}-v_{n-1})
  %@@x_{n-1}y_{n-1}-x_{n+1}y_{n+1})+
  +b(x_{n+1}y_{n}-x_{n}y_{n-1})\no\\
&&+c(2y_nx_{n+2}v_{n+1}-2x_ny_{n-2}v_{n-1}+x_n^2y^2_{n-1}-y
^2_nx^2_{n+1})=0 \label{identity2'}
  \eea

\proof At first, compute the first and second
difference in (\ref{identity1}) and (\ref{identity2}),
   \bea
\lefteqn{ \pl_n (\LR^{(n)}_1-\LR^{(n)}_2)_{n,n}}\no\\
 &=&
(\LR^{(n)}_1-\LR^{(n)}_2)_{n+1,n+1} -
 (\LR^{(n)}_1-\LR^{(n)}_2)_{n,n}
 + (c  L_1-aL_2)_{n+1,n+1} \label{proof1}\\&&\no\\
\lefteqn{ \pl^2_n
(v_{n-1}\LR^{(n)}_1-\LR^{(n)}_2)_{n,n-1}}\no\\
  &=&
 (v_{n+1}\LR_1^{(n)}-\LR_2^{(n)})_{n+2,n+1}
  +(v_{n-1}\LR_1^{(n)}-\LR_2^{(n)})_{n,n-1}
  -2(v_n\LR_1^{(n)}-\LR_2^{(n)})_{n+1,n}\no\\
 &&+ 2c\left(v_{n+1}( L_1)_{n+2,n+1}-
    v_{n}( L_1)_{n+1,n}\right)
    -2a \bigl(( L_2)_{n+2,n+1}-
         (L_2)_{n+1,n}\bigr)\no\\
 &=&
 (v_{n+1}\LR_1^{(n)}-\LR_2^{(n)})_{n+2,n+1}
  +(v_{n-1}\LR_1^{(n)}-\LR_2^{(n)})_{n,n-1}
  -2(v_n\LR_1^{(n)}-\LR_2^{(n)})_{n+1,n}\no\\
 &&+ 2\pl_n\left(
 c v_{n}(  L_1)_{n+1,n}
    -a (L_2)_{n+1,n}\right)\no\\
    \eea
In order to obtain the first identity, namely (\ref{identity1}),
form arbitrary linear combinations of the Virasoro equations
(\ref{Virasoro}),
 \bean
  \lefteqn{ \frac{1}{\tau_n^{\gamma+\vr}}
 \left( a\BV^{\gamma+\vr}_{-1,n}+b\BV^{\gamma+\vr}_{0,n}
  +c\BV^{\gamma+\vr}_{1,n}
   \right)\tau_n^{\gamma+\vr}}\\
 &=&\left\{
 \begin{array}{l}
 \displaystyle{
 \sum_{i\geq 1}\Bigl(
(a(i+1)t_{i+1}+bit_i+c(i-1)t_{i-1})\frac{\pl}{\pl
t_{i}}
 } \\
 \displaystyle{  -\left(a(i-1)s_{i-1}+bis_i+
 c(i+1)s_{i+1}\right)\frac{\pl}{\pl s_{i}}
   \Bigr)
   } \\
  \displaystyle{
  + c(n+\gamma)\frac{\pl}{\pl
  t_{1}}
   + a(n-\gamma)\frac{\pl}{\pl
  s_{1}}
  }
  \end{array} \right\}
  \log \tau^{\gamma+\vr}_n\\
&& +\vr \left( ( -a\frac{\pl}{\pl
  s_{1}}
  +
  c\frac{\pl}{\pl
  t_{1}} )
  \log \tau^{\gamma+\vr}_n +bn   \right)+b\gamma n
  \\&&\\
  &=&
   \sum_{i\geq 1}\left(\al_i(t)\frac{\pl}{\pl
  t_{i}}-\beta_i (s)\frac{\pl}{\pl
  s_{i}}\right) \log \tau^{\gamma+\vr}_n
  \\
  &&\\
  &&
  +\vr \left( ( c\frac{\pl}{\pl
  t_{1}}-a\frac{\pl}{\pl
  s_{1}}
   )
  \log \tau^{\gamma+\vr}_n +bn\right)
  +b\gamma n=
  0.
\eean
 Subtracting the $\vr=\pm 1$ contribution from
the $\vr=0$ contribution and omitting the lower index
$n$ in $\BV^{\gamma+\vr}_{k,n}$, leads to
\bea
 0&=& \frac{1}{\tau_n^{\gamma+\vr}}
 \left(a \BV^{\gamma+\vr}_{-1}+b\BV^{\gamma+\vr}_{0}
  +c\BV^{\gamma+\vr}_{1}
   \right)\tau_n^{\gamma+\vr}
   -
   \frac{1}{\tau_n^{\gamma}}
 \left( a\BV^{\gamma}_{-1}+b\BV^{\gamma}_{0}
  +c\BV^{\gamma}_{1}
   \right)\tau_n^{\gamma}\no\\
 &=&\sum_{i\geq 1}\left(\al_i(t)\frac{\pl}{\pl
  t_{i}}-\beta_i(s) \frac{\pl}{\pl
  s_{i}}\right)
  \log \frac{\tau^{\gamma+\vr}_n}{\tau^{\gamma}_n}+\vr \left(
   ( c\frac{\pl}{\pl
  t_{1}}-a\frac{\pl}{\pl
  s_{1}}
   )
  \log \tau^{\gamma+\vr}_n +bn   \right)\no\\
\label{intermediate}\eea
 Set, for brevity, $\tau_n=\tau_n^{\gamma}$ and
 $\tau_n^{\pm}=\tau_n^{\gamma+\vr},~
  \BV=\BV^{\gamma},~\BV^{\pm}=\BV^{\gamma+\vr}$, with
  $\vr=\pm 1$.

\medbreak

\noindent $\bullet$ {\bf CASE 1 and 2}: when not both
$a=c=0$, the terms $\frac{\pl}{\pl
  t_{1}}\log \tau^{\gamma+\vr}_n$ and $\frac{\pl}{\pl
  s_{1}}
  \log \tau^{\gamma+\vr}_n$ are present; they cannot
  be easily expressed in terms of $x_n$ and $y_n$. But
 adding the $+$contribution to the $-$contribution
 eliminates those terms:
    \bean
  \lefteqn{0=\frac{x_ny_n}{v_n}\left\{ \begin{array}{l}
 \displaystyle{ \frac{1}{\tau_n^{+}}
 \left(a \BV^{+}_{-1}+b\BV^{+}_{0}
  +c\BV^{+}_{1}
   \right)\tau_n^{+}
   +
   \frac{1}{\tau_n^{-}}
 \left(a \BV^{-}_{-1}+b\BV^{-}_{0}
  +c\BV^{-}_{1}
   \right)\tau_n^{-} } \\
  \hspace{2cm}\displaystyle{ -
   \frac{2}{\tau_n^{}}
 \left( a\BV^{}_{-1}+b\BV^{}_{0}
  +c\BV^{}_{1}
   \right)\tau_n^{}}\end{array}\right\}}\\
 &=&\frac{x_ny_n}{v_n}
 \left(\sum_{i\geq 1}\left(\al_i(t)\frac{\pl}{\pl
  t_{i}}-\beta_i(s) \frac{\pl}{\pl
  s_{i}}\right)\log x_ny_n
 +\left(
  c\frac{\pl}{\pl
  t_{1}}  -a\frac{\pl}{\pl
 s_{1}}
     \right)\log \frac{x_n}{y_n}\right).
\\
&=&\frac{x_ny_n}{v_n}
 \left( \sum_{i\geq 1}
  (\alpha_i  \frac{\pl}{\pl t_{i}}
  -\beta_i  \frac{\pl}{\pl s_{i}}  )
  (\log x_n+\log y_n)
 + (c  \frac{\pl}{\pl t_{1}}
  -a \frac{\pl}{\pl s_{1}} )
  (\log x_n-\log y_n) \right )
  \\ &&
\\&=&
   (\LR^{(n)}_1-\LR^{(n)}_2+aL_2-cL_1)_{nn}-
  (\LR^{(n)}_1-\LR^{(n)}_2-aL_2+cL_1)_{n+1,n+1},
 \\&&\hspace{2cm}~~ \mbox{using (\ref{tbasicvariables}), (\ref{tderivative})
  and the definition (\ref{LR}) of $\LR_i^{(n)}$, }\\&=&
  -\pl_n (\LR_1^{(n)}
   -\LR_2^{(n)})_{n,n}
  +(aL_2-c L_1)_{nn}\eean
   using (\ref{proof1}). This establishes the first
   relation, namely (\ref{identity1}).

To prove the second relation (\ref{identity2}), we
take the $t_1$-derivative of the first relation.
 Using
  $$
\begin{array}{lll}
\displaystyle{\frac{\pl\al_i}{\pl t_1}}&=0&\mbox{for~}
i\geq 3\\ &=c&\mbox{for~} i=2
 \\
  &=b&\mbox{for~} i=1
\\
 \displaystyle{\frac{\pl\beta_i}{\pl t_1}}&=0&\mbox{for~} i\geq 1
 \end{array}
  $$
 we obtain
  $$ \frac{\pl(\LR_1-\LR_2)}{\pl
t_1}=\sum_{i\geq 1} \al_i(t)\frac{\pl L^i_1}{\pl t_1}+
\sum_{i\geq 1} \beta_i(t)\frac{\pl L^i_2}{\pl t_1}+c
L^2_1+b  L_1 $$

Using the above and
  \bean
   ( L^2_1)_{n,n}&=&x^2_ny^2_{n-1}-x_ny_{n-2}
v_{n-1}-x_{n+1}y_{n-1}v_n  \\
(
L^2_2)_{n,n}&=&y^2_nx^2_{n-1}-y_nx_{n-2}
 v_{n-1}-y_{n+1}x_{n-1}v_n
 \\
 (
L^2_2)_{n,n-1}&=&-v_{n-1}(x_{n-1}y_n+x_{n-2}y_{n-1})
 \eean
 one computes, using (\ref{LR}) and Lemma 1.3,
\bean
 0&=&
 \frac{\pl}{\pl
t_1}(\LR_1-\LR_2+aL_2-c L_1)_{n,n}-\frac{\pl}{\pl
t_1}(\LR_1-\LR_2-aL_2+c  L_1)_{n+1,n+1}
\\
&=&
  \left(\frac{\pl(\LR_1-\LR_2)}{\pl
t_1}\right)_{n,n}- \left(\frac{\pl(\LR_1-\LR_2)}{\pl
t_1}\right)_{n+1,n+1}\\ &&
 \hspace{1cm}
 +a\left(\left(\frac{\pl
L_2}{\pl t_1}\right)_{n,n}+\left(\frac{\pl L_2}{\pl
t_1}\right)_{n+1,n+1}\right)\\&&\hspace{1cm}
  -c
\left(\left(\frac{\pl   L_1}{\pl
t_1}\right)_{n,n}+\left(\frac{\pl  L_1}{\pl
t_1}\right)_{n+1,n+1}\right) \\
 &=&\sum^{}_{i\geq 1}\al_i(t)
%  \\
% &&
 \left(2v_n(
L_1^i)_{n+1,n}-v_{n-1}( L^i_1)_{n,n-1}-v_{n+1}(
L^i_1)_{n+2,n+1}\right)\\ &
&+\sum^{}_{i\geq 1}\beta_i(t)
 \left(2(L_2^i)_{n+1,n}-(L_2^i)_{n,n-1}-(L_2^i)_{n+2,
n+1}\right)\\ & &+c\left(( L^2_1)_{nn}-(
L^2_1)_{n+1,n+1}-v_{n+1}( L_1)_{n+2,n+1}+ v_{n-1}(
L_1)_{n,n-1}\right)\\ &
&+a\left((L_2)_{n+2,n+1}-(L_2)_{n,n-1}\right)
 +b\left((
L_1)_{nn}-( L_1)_{n+1,n+1}\right)\\
&=&2v_{n}(\LR_1)_{n+1,n}-v_{n-1}(\LR_1)_{n,n-1}-
 v_{n+1}(\LR_1)_{n+2,n+1}\\ &
&-2(\LR_2)_{n+1,n}+(\LR_2)_{n,n-1}+(\LR_2)_{n+2,n+1}\\
& &+c\left(( L^2_1)_{nn}-( L^2_1)_{n+1,n+1}-v_{n+1}(
L_1)_{n+2,n+1}+ v_{n-1}( L_1)_{n,n-1}\right)\\ &
&+a\left((L_2)_{n+2,n+1}-(L_2)_{n,n-1}\right)
 +b\left((
L_1)_{nn}-( L_1)_{n+1,n+1}\right)\\
  &\stackrel{*}{=}& 2(v_n\LR_1-\LR_2)_{n+1,n}
  -(v_{n-1}\LR_1-\LR_2)_{n,n-1}
  -(v_{n+1}\LR_1-\LR_2)_{n+2,n+1}\\
  && +a(v_{n+1}-v_{n-1})+
  b(x_{n+1}y_{n}-x_{n}y_{n-1})\\
&&+c(2y_nx_{n+2}v_{n+1}-2x_ny_{n-2}v_{n-1}+x_n^2y^2_{n-1}-y
^2_nx^2_{n+1})   \\
  &=& 2(v_n\LR_1-\LR_2)_{n+1,n}
  -(v_{n-1}\LR_1-\LR_2)_{n,n-1}
  -(v_{n+1}\LR_1-\LR_2)_{n+2,n+1}\\
  &&
  +\pl_n\left(
 \begin{array}{l}
  a\left((L_2)_{n,n-1} + (L_2)_{n+1,n}\right) \\
  -\left(c( L_1^2)_{nn}+b( L_1)_{nn}\right) \\
  -c\left( v_n( L_1)_{n+1,n}+
   v_{n-1}( L_1)_{n,n-1}\right)
   \end{array} \right)
\\
 &\stackrel{**}{=}&
 -\pl_n^2(v_{n-1}\LR_1-\LR_2)_{n,n-1}
 +2\pl_n \left( cv_n( L_1)_{n+1,n}
 -a(L_2)_{n+1,n}
   \right)
    \\&&
  +\pl_n\left(
 \begin{array}{l}
  a\left((L_2)_{n,n-1} + (L_2)_{n+1,n}\right) \\
  -\left(c( L_1^2)_{nn}+b( L_1)_{nn}\right) \\
  -c\left( v_n( L_1)_{n+1,n}+
   v_{n-1}( L_1)_{n,n-1}\right)
   \end{array} \right)
\eean
\bean
 &=&
 -\pl_n^2(v_{n-1}\LR_1-\LR_2)_{n,n-1}
    \\&&
  +\pl_n\left(
 \begin{array}{l}
  a\left((L_2)_{n,n-1} - (L_2)_{n+1,n}\right) \\
  -\left(c( L_1^2)_{nn}+b( L_1)_{nn}\right) \\
  +c\left( v_n( L_1)_{n+1,n}-
   v_{n-1}( L_1)_{n,n-1}\right)
   \end{array} \right)
  \\
  &=&
 -\pl_n^2(v_{n-1}\LR_1^{(n)}- \LR_2^{(n)}
 +aL_2-cv_{n-1}L_1)_{n,n-1}
  \\&& -\pl_n \left(c( L_1^2)_{nn}+b( L_1)_{nn}\right)
  \\
  &=&
 -\pl_n^2(v_{n-1}\LR_1^{(n-1)}-\LR_2^{(n-1)}
 )_{n,n-1}
  -\pl_n \left(c( L_1^2)_{nn}+b( L_1)_{nn}\right).
 \\
 &=&
 -\pl_n\left(\pl_n(v_{n-1}\LR_1^{(n-1)}-\LR_2^{(n-1)}
 )_{n,n-1}
  +\bigl(c( L_1^2)_{nn}+b( L_1)_{nn}\bigr)
  \right),\eean
using (\ref{proof2}) in $\stackrel{**}{=}$, ending the proof of
identity (\ref{identity2}). Equality $\stackrel{*}{=}$ leads to
the expression in the remark after the statement of the Theorem.

\medbreak

\noindent $\bullet$ {\bf CASE 3}: when both $a=c=0$,
the terms $\frac{\pl}{\pl
  t_{1}}\log \tau^{\gamma+\vr}_n$ and $\frac{\pl}{\pl
  s_{1}}
  \log \tau^{\gamma+\vr}_n$ are absent in
  (\ref{intermediate}). So, using again
  (\ref{tderivative}), setting $\al_i(t)=it_i$,
  $\beta_i(s)=is_i$ and $b=1$, leads to the polynomials
\bean
 0&=& -x_n
 \left(\sum_{i\geq 1}\left(\al_i(t)\frac{\pl}{\pl
  t_{i}}-\beta_i(s) \frac{\pl}{\pl
  s_{i}}\right)
  \log x_n+bn\right)\\
  &=&
   \frac{v_n}{y_n}\sum_{i\geq 1}^{}\left\{\al_i
  \left(( L_1^{i})_{n+1,n+1}-
   ( L_1^{i-1})_{n+1,n}\right)
    -
  \beta_i
  \left(  (L_2^i)_{nn} -  (L_2^{i-1})_{n,n+1}\right)
  \right\}
  \\
 &&\hspace{9cm}-n{x_n}{}
 \eean

\bea
 0&=&
  y_n
 \left(\sum_{i\geq 1}\left(\al_i(t)\frac{\pl}{\pl
  t_{i}}-\beta_i(s) \frac{\pl}{\pl
  s_{i}}\right)
  \log y_n-bn\right)%= R_2^{(3)}(n)
\no\\ &=&
 \frac{v_n}{x_n}\sum_{i\geq 1}^{} \left\{
  \al_i\left(( L_1^i)_{nn}-( L_1^{i-1})_{n+1,n}
  \right)   -\beta_i
 \left((L_2^i)_{n+1,n+1}-(L_2^{i-1})_{n,n+1}\right)
 \right\}
  \no\\
&&\hspace{9cm}-n y_n , \label{case3}
 \eea
 ending the proof of
Theorem 2.2. \qed

\subsection{Proof of main Theorem}

 {\medskip\noindent{\it Proof of Theorem 0.1:\/} }
Remember the locus, defined in (\ref{locus}),
 \bea \LR&=&\left\{
\begin{array}{l}
it_i=it_i^{(0)}:=\left\{\begin{array}{l}
u_i-(\gamma'_1d_1^i+\gamma'_2d_2^i),\mbox{~for~}1\leq
i\leq N_1\\
 \\
-(\gamma'_1d_1^i+\gamma'_2d_2^i),\mbox{~for~}N_1+1\leq
i<\infty
\end{array}\right.\\
 \\
is_i=is_i^{(0)}:=\left\{\begin{array}{l} -u_{-i}+
(\gamma''_1d_1^{-i}+\gamma''_2d_2^{-i}),\mbox{~for~}1\leq
i\leq N_2\\
 \\
(\gamma''_1d_1^{-i}+\gamma''_2d_2^{-i}),\mbox{~for~}N_2+1\leq
i<\infty
\end{array}\right.
\end{array}\right\}\nonumber\\ \label{locus1}
\eea
 From (\ref{relations}) in Lemma 2.1, for all $i\geq 1$,
   \bean
\alpha_i(t^{(0)})&:=&
 au_{i+1} +bu_{i} +cu_{i-1}
 +c(n+\gamma_1^{\prime}+
  \gamma_2^{\prime}+\gamma)\dt_{i1}
  \\
\beta_i(s^{(0)})&:=&
 -au_{-i+1} -bu_{-i} -cu_{-i-1}
 -a(n+\gamma_1^{\prime\prime}+
  \gamma_2^{\prime\prime}-\gamma)\dt_{i1}
.\eean
 Then the locus (\ref{locus1}) can equally be described
 by
 \be\LR=\left\{
  \begin{array}{l}
  ~\mbox{all}~\al_{i}(t)=0,
 ~\mbox{for $i\geq N_1+2$ and all}~
~\beta_{i}(t)=0,
 ~\mbox{for $i\geq N_2+2$,}~\\
  \al_{i}(t)=au_{i+1}+bu_i+c\bigl(
 u_{i-1}+ \dt_{i1}(n+\gamma_1^{\prime}+
  \gamma_2^{\prime}+\gamma)\bigr) ~~ \mbox{and}~~ \\
 \beta_{i}(s)= -c
 u_{-i-1}-bu_{-i}-a\bigl(u_{-i+1}+
 \dt_{i1}(n +\gamma_1^{\prime\prime}+
  \gamma_2^{\prime\prime}-\gamma)\bigr),\\
  \hspace{8cm}~~\mbox{otherwise}
 \end{array}
 \right\}
  ,\label{locus2}\ee
 The $\LR_i$-matrices (\ref{LR}) now and only now are finite sums
 and so have the form,
 setting
 $u_0=u_{N_1+1}=u_{N_1+2}=u_{-N_2-1}=u_{-N_2-2}=0$,
 \bean
  \LR_1&=&\sum_1^{N_1+1} \al_i(t^{(0)})  L_1^i
   \\
   &=& \sum_1^{N_1+1}
   \left( au_{i+1}+bu_i+c u_{i-1}\right)
    L_1^i+
    c\bigl(
 n+\gamma_1^{\prime}+
  \gamma_2^{\prime}+\gamma\bigr)  L_1\no\\
  &=&
  \left(aI+b L_1+c  L_1^2\right)
  \sum_1^{N_1} u_i  L_1^{i-1}
  + c \bigl( n+\gamma_1^{\prime}+
  \gamma_2^{\prime}+\gamma\bigr)  L_1-au_1I\no\\
&=&
  \left(aI+b L_1+c  L_1^2\right)
  P_1^{\prime}(  L_1)
  + c \bigl( n+\gamma_1^{\prime}+
  \gamma_2^{\prime}+\gamma\bigr)  L_1-au_1I\no\\
  \eean \bean
  \LR_2&=&-\sum_1^{N_2+1}\beta_i (s^{(0)})
   L_2^i
   \\
   &=& \sum_1^{N_2+1}
   \left( cu_{-i-1}+bu_{-i}+a u_{-i+1}\right)
    L_2^i+
    a\bigl(
 n+\gamma_1^{\prime\prime}+
  \gamma_2^{\prime\prime}-\gamma\bigr)  L_2\no\\
  &=&
  \left(cI+b L_2+ a  L_2^2\right)
  \sum_1^{N_2} u_{-i}  L_2^{i-1}
  + a \bigl( n+\gamma_1^{\prime\prime}+
  \gamma_2^{\prime\prime}-\gamma\bigr)  L_2-cu_{-1}I\no\\
&=&
  \left(cI+b L_2+a  L_2^2\right)
  P_2^{\prime}(  L_2)
  + a \bigl( n+\gamma_1^{\prime\prime}+
  \gamma_2^{\prime\prime}-\gamma\bigr)  L_2-cu_{-1}I\no
   %\label{LR}
   ,\eean
which are precisely the expressions $\LR^{(n)}_i$,
introduced in (\ref{L-matrices}), in the introduction,
 but modulo the identity pieces.
 Still, the identities (\ref{identity1}),
(\ref{identity2}) and (\ref{identity3}) remain valid,
upon evaluating along the locus $\LR$; i.e., with
$\al_i(t^{(0)})$ and $\beta_i(s^{(0)})$ as in
(\ref{locus2}) and $t$ and $s$ replaced by $t^{(0)}$
and $s^{(0)}$ in the variables $x_n(t,s)$ and
$y_n(t,s)$.

\noindent $\bullet$ CASE 1 and 2: not both $a=c=0$.
  Thus, the first identity (\ref{identity1}) holds and
  the
second identity (\ref{identity2}) expresses the fact
that a difference $\pl_n$ of an expression vanishes;
therefore the expression equals that same expression
at the origin. This ends the proof of identities
(\ref{recurrence1}) and (\ref{recurrence2}) in Theorem
0.1, upon observing the identity pieces in $\LR_1$ and
$\LR_2$ above make no contribution.

\noindent $\bullet$ CASE 3: both $a=c=0$ and $b=1$.
From (\ref{locus}) or (\ref{locus1}), setting $a=c=0$
and $b=1$, we have
 $$
\al_i(t^{(0)})=u_i~~~\mbox{and}~~~\beta_i(t^{(0)})=-u_{-i}
 $$
  Remember from (0.0.2), $ P_1(z):=\sum_1^{N_1} \frac{u_iz^i}{i}
  ~~\mbox{and}
  ~~
  P_2(z):=\sum_1^{N_2} \frac{u_{-i}z^{i}}{i},
 $ and so we have
 \bean
 \sum_{i\geq 1}^{}  \al_i (t^{(0)})
 L_1^{i-1}&=&\sum_{i\geq 1}^{}  u_i
 L_1^{i-1}=P_1^{\prime}(L_1)\\
 \sum_{i\geq 1}^{}  \beta_i (t^{(0)})
 L_2^{i-1}&=&-\sum_{i\geq 1}^{}  u_{-i}
 L_2^{i-1}=-P_2^{\prime}(L_2),%~~~\mbox{and thus}~~~
 \eean
and so (\ref{identity3}) leads immediately to (\ref{recurrence3}),
ending the proof of Theorem 0.1.
 \qed

%\newpage

\section{Invariant manifolds for the first Toeplitz flow}

{%\medskip
\noindent{\it Proof of Theorem 0.4:\/} }
The weight here is
 $$
 \rho(z):=
 z^{\gamma} e^{P_1(z)+P_2(z^{-1})},
 $$
corresponding to case 3 of
 Theorem 0.1.
 From the latter, the variables $x_n,y_n$ satisfy
 recurrence relations
 \bea
 \left\{
 \begin{array}{l}
 \Gamma_n(x,y)%=R^{(3)}_1(x,y)_n
 :=
  \displaystyle{\frac{v_n}{y_n}}
  \left(
  \begin{array}{l}
  -\left( L_1 P_1^{\prime} (
L_1)\right)_{n+1,n+1}
  - \left(  L_2 P_2^{\prime} (
L_2)\right)_{n,n} \\  \\
 + (P_1^{\prime}( L_1))_{n+1,n}
  + (P_2^{\prime}( L_2))_{n,n+1}
 \\
 \end{array}\right)  +nx_n=0
  \\ \\
%\hspace{9cm}-n\frac{x_ny_n}{v_n}=0\no\\\no \\
%
\tilde\Gamma_n(x,y)%= R^{(3)}_2(x,y)_n
  :=
  \displaystyle{\frac{v_n}{x_n}  }
\left(
\begin{array}{l}
  -\left( L_1 P_1^{\prime} (
L_1)\right)_{n,n}
 -\left(  L_2 P_2^{\prime} (
L_2)\right)_{n+1,n+1} \\  \\
 + (P_1^{\prime}(
L_1))_{n+1,n}
  + (P_2^{\prime}( L_2))_{n,n+1}
  \end{array}\right) +ny_n=0,
  \no\\
%\hspace{9cm}-n\frac{x_ny_n}{v_n}=0 . \\
%
\end{array} \right.\\
\label{recurrence31}
   \eea
 which by virtue of (\ref{case3}) and the the nature of
 the locus
 (\ref{locus1}), can be written
 \bea
  \Gamma_n
  %R^{(3)}_1(n)
   &=&%-\frac{y_n}{v_n}
  {\cal V}_0 x_n+nx_n \no\\
 \tilde\Gamma_n
 %R^{(3)}_2(n)
 &=&%\frac{x_n}{v_n}
 -{\cal V}_0 y_n+ny_n,
 \label{Gammas}\eea
  where
 $$
 \VR_0:= \sum_{i\geq 1} \left(u_i%\al_i(t)
\frac{\pl}{\pl t_i}
 +u_{-i}%-\beta_i(s)
  \frac{\pl}{\pl s_i}\right),
 $$
in terms of the Toeplitz vector fields.
Recall from section 6 (Appendix 2) the
form of the first vector field:
($v_n:=1-x_ny_n$)
 \bean
  \frac{\pl x_n}{\pl
  \left\{\displaystyle{{t_1}\atop {s_1}}\right\}}
  =v_nx_{n\pm 1}  &~~~~~&
\frac{\pl y_n}{\pl  \left\{\displaystyle{{t_1}\atop
{s_1}}\right\}}=-v_ny_{n\mp 1}.
% \\&&
% \hspace{4cm}(\mbox{\bf Toeplitz
%Lattice}) \\ \frac{\pl x_k}{\pl s_1}=v_kx_{k-1}
%&~~~~~& \frac{\pl y_k}{\pl s_1}=-v_ky_{k+1}  .
% \\
 \eean
Also in statement of Theorem 0.4, we assume the $u_i$, appearing in the polynomials $P_1(z)$
and $P_2(z)$, flow according to
  \be
  \frac{\pl u_k}{\pl t_1}=\dt_{k,1}
  ~~~~~~~~~~~~\frac{\pl u_k}{\pl s_1}=-\dt_{k,-1}.
  \label{u-flow}
  \ee
Noticing that, from (\ref{u-flow}),
 $$\left[\frac{\pl }{\pl t_1},\VR_0\right]=\frac{\pl }{\pl
 t_1}~~,~~
  \left[\frac{\pl }{\pl s_1},\VR_0\right]=-\frac{\pl }{\pl
 s_1},
 $$
 we compute
\bea \frac{\pl}{\pl t_1}\Gamma_n
 &=&
\VR_0\frac{\pl x_n}{\pl t_1}
 +\frac{\pl x_n}{\pl t_1}
  +n\frac{\pl x_n}{\pl t_1}\no\\
  &=&
 \VR_0 (v_nx_{n+1})+(n+1)v_nx_{n+1}\no\\
 &=&
  v_n\left(\VR_0 x_{n+1}+(n+1)x_{n+1}\right)
   +x_{n+1}\VR_0(v_n)\no\\
 &{=}&
  v_n\Gamma_{n+1}-x_{n+1}\left(x_n \VR_0(y_n)+
   y_n \VR_0(x_n)\right)\no\\
    &=&
  v_n\Gamma_{n+1}-x_{n+1}\left(x_n (\VR_0(y_n)-ny_n)+
   y_n (\VR_0(x_n)+nx_n)\right)\no\\
   &=&
  v_n\Gamma_{n+1}+x_{n+1}\left(x_n \tilde\Gamma_n-
   y_n \Gamma_n\right),
  \label{Gamma-flow1}\eea
Similarly, one shows
\bea
 \frac{\pl}{\pl s_1}\Gamma_n
 &=&
\VR_0\frac{\pl x_n}{\pl s_1}
 -\frac{\pl x_n}{\pl s_1}
  +n\frac{\pl x_n}{\pl s_1}\no\\
  &=&
 \VR_0 (v_nx_{n-1})+(n-1)v_nx_{n-1}\no\\
 &=&
  v_n\Gamma_{n-1}+  x_{n-1}\left(x_n \tilde\Gamma_n-
   y_n \Gamma_n\right)\label{Gamma-flow2}
\eea
and so by the duality $\tilde{}$
\bea
   \frac{\pl}{\pl t_1}\tilde\Gamma_n
 &=&
  -v_n\tilde\Gamma_{n-1}+ y_{n-1}\left(
   x_n \tilde\Gamma_n-
   y_n \Gamma_n\right)
  \no\\&&\no\\
   \frac{\pl}{\pl s_1}\tilde\Gamma_n
 &=&
  -v_n\tilde\Gamma_{n+1}+y_{n+1}\left(x_n
  \tilde\Gamma_n   -
   y_n \Gamma_n\right),
    \label{Gamma-flow3}
 \eea
thus establishing (\ref{gamma-ode1}). Setting as
in (\ref{M-locus}),
 $$
    {\frak
M}%^{(k)}
 :=\bigcap_{n\geq 0} \left \{ (x_k,y_k)_{k\geq
0}~,~~\mbox{such that}~~
\Gamma_n(x,y)=0%R^{(k)}_1(x,y)_n=0,~R^{(k)}_2(x,y)_n=0
 \mbox{ and }\tilde\Gamma_n(x,y)=0\right\}
, %\label{M-locus}
  $$
 the differential equations (\ref{Gamma-flow1}), (\ref{Gamma-flow2}) and (\ref{Gamma-flow3}) imply
 at once that,  along the locus ${\frak M}$, defined
 in (\ref{M-locus}),
  $$
 \left.\frac{\pl}{\pl t_1}\Gamma_n\right|
 _{\frak M}=
  \left.\frac{\pl}{\pl t_1}\tilde\Gamma_n\right|
 _{\frak M}
 =\left.\frac{\pl}{\pl s_1}\Gamma_n\right|
 _{\frak M}=
  \left.\frac{\pl}{\pl s_1}\tilde\Gamma_n\right|
 _{\frak M}=0,
  $$
showing the locus $\frak M$ is an invariant manifold for these
flows. This ends the proof of Theorem 0.4. \qed

{\medskip
\noindent{\it Proof of Corollary 0.5:\/} }
In the self-dual case,
$$\Gamma_n=\tilde\Gamma_n,  $$
since in (\ref{Gammas}) all $u_n=u_{-n},~x_n=y_n$ and so $\pl x_n / \pl
t_i= -\pl y_n / \pl s_i$, $\pl x_n / \pl
s_i= -\pl y_n / \pl t_i$ for all $n,i$.

From the differential equations
(\ref{Gamma-flow1}) and (\ref{Gamma-flow2}),
it follows that
 $$\frac{\pl}{\pl t_1}\Gamma_n=v_n\Gamma_{n+1}~~~~\mbox{and}
  ~~~~ \frac{\pl}{\pl s_1}\Gamma_n=v_n\Gamma_{n-1}, $$
 and so for
$\frac{\pl}{\pl t}=\frac{\pl}{\pl t_1}-\frac{\pl}{\pl
s_1},$

$$ \frac{\pl \Gamma_n}{\pl t}=v_n (\Gamma_{n+1}-
 \Gamma_{n-1})$$
establishing equation (\ref{Gamma-ode-intro}).
 Therefore, along the locus $\frak N$, defined in
 (\ref{N-locus}),
  $$
 \left.\frac{\pl \Gamma_n}{\pl t} \right|
 _{\frak N}=0
 $$
 and so the locus $\frak N$ is invariant with respect
 to the $\pl /\pl t$ vector field, ending the proof of
 Corollary 0.5. \qed

%\newpage

\section{Rational relations for special weights}

\subsection{Weight $ e^{t(z+z^{-1})}
$}
This weight comes up by considering the uniform probability $P$ on the group
 $ S_k$ of permutations $\pi_k$ and
\be
L(\pi_k) =  \mbox{ length of the longest (strictly) increasing
subsequence of $\pi_k$ }.
 \label{longest}\ee
Then, according to Gessel \cite{Gessel}, the
generating function below can be expressed as
the determinant of a Toeplitz matrix and thus as a unitary matrix integral:
  \bea
\sum^{\iy}_{k=0}\frac{t^{2k}}{k!}P(L(\pi_k)\leq n)
 &=&E_{U(n)}e^{{t}\Tr (M+\bar M)}\\
&=&
\frac{1}{n!}
\int_{(S^1)^{n}}|\Dt_n(z)|^{2}
 \prod_{k=1}^n
\left(%z_k^{\vr}
e^{t(z_k+\bar z_k)}
 \frac{dz_k}{2\pi i z_k}\right).\nonumber
 \eea
Then
\bean
x_n
  &=&
  (-1)^n\frac{E_{U(n)}(\det M ) e^{{t}\Tr (M+\bar M)}}
  {E_{U(n)}e^{{t}\Tr (M+\bar M)}}
  \\
  &=& (-1)^n
  \frac{\int_{(S^1)^{n}}|\Dt_n(z)|^{2}
 \prod_{k=1}^n
\left( z_k e^{t(z_k+\bar z_k)}
 \frac{dz_k}{2\pi i z_k}\right)}
 {\int_{(S^1)^{n}}|\Dt_n(z)|^{2}
 \prod_{k=1}^n
\left(   e^{t(z_k+\bar z_k)}
 \frac{dz_k}{2\pi i z_k}\right)}
  \eean

This weight is a special case of the self-dual
 weight of Corollary 0.3, which lead to the relation
 (\ref{recurrence3self}). Thus we find
 \bean
0&=&nx_n-\frac{v_n}{x_n}\left(
 t(L_1)_{n+1,n+1}+t(L_1)_{nn} \right)\\
  &=&nx_n-\frac{v_n}{x_n}\left(
 -tx_{n+1}x_{n}-t x_{n}x_{n-1} \right)
  \eean
   yielding the $3$-step relation, found by Borodin
    \cite{B},
 with highest (respectively, lowest)
 terms doubly (respectively, simply) underlined,
   \be n{x_n}+t(1-x_n^2)(\underline{\underline{x_{n+1}}}+
\underline{x_{n-1}})=0 .
 \label{McMillan}\ee

 It is interesting to point out that this map (\ref{McMillan})
  is the simplest instance of a family of
  area-preserving maps of the
 plane, having an invariant, as found by
 McMillan \cite{McMillan}, and extended by Suris \cite{Suris} to maps of the
 form $\pl_n^2x(n)=f(x(n))$, having an analytic invariant
 of two variables $\Phi(y,z)$, i.e.,
 $$ \Phi(x_{n+1}, x_n)=\Phi(x_n,x_{n-1}).$$
 The invariant in the case of the maps
 (\ref{McMillan}) is
 $$ \left(1-y^{2}\right)\left(1-z^{2}
 \right)+ayz ,~~\mbox{with} ~a=-\frac{n}{t}.$$
 For more on this matter, see the review by B. Grammaticos,
  F. Nijhoff, A. Ramani \cite{Nijhoff}.

\subsection{Weight $ e^{t(z+z^{-1})+s(z^2+z^{-2})}
$}

Consider instead the subgroups of odd permutations, with $2^k k!$
elements
$$
S^{\mbox{\tiny odd}}_{2k} = \left\{
\begin{array}{l}
\pi_{2k} \in S_{2k}, \pi_{2k}: (-k, \ldots, -1, 1, \ldots, k)
\circlearrowleft\\
\mbox{with }\pi_{2k} (-j)=-\pi_{2k} (j), \mbox{ for all }j
\end{array}
\right\} \subset S_{2k}
$$
$$
S^{\mbox{\tiny odd}}_{2k+1} = \left\{
\begin{array}{l}
\pi_{2k+1} \in S_{2k+1}, \pi_{2k}: (-k, \ldots, -1,0, 1, \ldots,
k)
\circlearrowleft\\
\mbox{with }\pi_{2k+1} (-j)=-\pi_{2k+1} (j), \mbox{ for all }j
\end{array}
\right\} \subset S_{2k}
$$
Then, according to Rains \cite{Rains} and Tracy-Widom \cite{TW1},
the following generating functions, again involving the length of
the longest necessary sequence, are related to matrix integrals:

\begin{eqnarray*}
\sum_0^{\infty} \frac{(\sqrt{2}~ s)^{2k}}{k!} P (L(\pi_{2k})\leq
n)
&=& E_{U(n)} e^{s Tr(M^2+\bar M^2)}\\
&=& \frac{1}{n!} \int_{(S')^n} |\Delta_n(z) |^2
 \prod_{k=1}^n \left(
e^{s(z_k^2+z_k^{-2})} \frac{dz_k}{2\pi z_k}\right)
\end{eqnarray*}
\bean
 \lefteqn{\sum_0^{\iy} \frac{(\sqrt{2}~ s)^{2k}}{k!}
 P (L(\pi_{2k+1})\leq n)
 }\\
 &=&
  \frac{1}{4} \frac{\pl}{\pl t^2} \left. \left(
E_{U(n)} e^{Tr (t(M+\bar M)+s(M^2+\bar M^2)} +E_{U(n)}
e^{Tr(t(M+\bar M)-s(M^2+\bar M^2))} \right)\right|_{t=0}
 \eean

The weight $e^{t(z+z^{-1})+s(z^2+z^{-2})}$ is a special case of
the one in corollary 0.3, thus leading to (0.0.17).  So, we find
$$
x_n=(-1)^n \frac{E_{U(n)} (\det M) e^{t Tr(M+\bar M)+s Tr(M^2+\bar
M^2)}}{E_{U(n)}e^{t Tr(M+\bar M)+s Tr(M^2+\bar M^2)} }
$$
satisfies a $5$-step relation,
 with highest (respectively, lowest)
 terms doubly (respectively, simply) underlined,
\be
 0= {nx_n}+t{v_n}(x_{n-1}+x_{n+1})+2s{v_n}\left(  \underline{\underline{x_{n+2}}}v_{n+1}+
  \underline{x_{n-2}}v_{n-1}
 -x_n(x_{n+1}+x_{n-1})^2  \right).
\label{5-step}
 \ee
  Also here the map has a polynomial invariant
  $$
  \Phi(x,y,z,u)= nyz-(1-y^2)(1-z^2)
   \Bigl( t+2s(x(u-y)-z(u+y))\Bigr).
  $$
 So we have for all $n$,
 $$
  \Phi(x_{n-1}, x_n, x_{n+1}, x_{n+2})=
  \Phi(x_{n}, x_{n+1}, x_{n+2}, x_{n+3}).
  $$

%\newpage

\subsection{Weight $ (1+{z} )^{\al} e^{-sz^{-1}}
$}

Let $P$ be the uniform probability on
$$
 S_{k,\al}=\{\mbox{words $\pi_k$ of length $k$ from an alphabet
of $\al$ letters}\}
. $$
Then, if $L$ denotes the same as in (\ref{longest}),
but without `strictly' , we have, according to Tracy-Widom
 \cite{TW} (see also \cite{AvM1}),
$$
\sum_{k=0}^{\iy}
 \frac{(\al s)^{k}}{k!}
P\left\{ \pi_k \in S_{k,\al} ~ \bigl|~
L(\pi_k)\leq n \right\}=
 E_{U(n)}~ \det (I+M)^{\al} e^{-s~ tr \bar M}
 .$$
For this weight, $P_1(z)=0$ and $P_2(z)=-sz$, so that $
N_1=0,~N_2=1,~u_{-1}=-s,\mbox{and all other }u_i=0.$
 Also
  $  \gamma_1^{\prime}=\al,~~
\gamma=\gamma_2^{\prime}=\gamma_1^{\prime\prime}=
\gamma_2^{\prime\prime}=0 $. This is a special case of
Case 2 of Theorems 0.1 and 0.2.
 Hence, we choose
$$a=0, ~~b=c=1,$$
for which one computes
  \bean \LR_1^{(n)}&=&  (n+\al) L_1  \\
      \LR_2^{(n)}&=& s(I+L_2),
\eean
and so (\ref{recurrence1}) and (\ref{recurrence2}) become
 \bean
&& \pl_n((n+\al)L_1+sL_2)_{nn}+c(L_1)_{nn}=0 \\
&&\pl_n((n-1+\al)v_{n-1}L_1+sL_2
 )_{n,n-1}
  +\bigl( L_1^2+ L_1\bigr)_{n,n}=\mbox{same}\Bigl
  |_{n=1}
 \eean
%
%\newpage
%
 Thus spelled out,
 the variables
$$
 {x_n \atop y_n}
=(-1)^n\frac
 { E_{U(n)}~ (\det M)^{\pm 1}\det (I+M)^{\al}
   e^{-s~ tr \bar M}}
   { E_{U(n)}~ \det (I+M)^{\al}
   e^{-s~ tr \bar M}}
 $$
satisfy a $3$-step and a
 $4$-step relation, linear in $x_{n+1}$ and $y_{n+1}$:
\bean
  - (n+\al+1)\underline{\underline{x_{n+1}}}y_{n}
  -s x_{n}\underline{\underline{y_{n+1}}}
  +(n+\al-1) x_ny_{n-1}
  +s x_{n-1}y_{n}
 =0
  \eean
  \bean
&&\hspace{-1cm}
-v_n((n+\al+1)\underline{\underline{x_{n+1}}}y_{n-1}
-s)+
  v_{n-1}((n+\al-2)x_{n}\underline{y_{n-2}}-s)
  \\&&~~~~~~~~~
  +x_n y_{n-1}(x_ny_{n-1}  -1)
  =v_1( s-(2+\al)x_{2})+x_1(x_1-1).
   \eean

%\newpage

\subsection{Weight $ \left(1-{\xi}{z} \right)^{\al}
   \left(1-{\xi}{z^{-1}}  \right)^{\beta}
$}

This weight, considered by Borodin \cite{B} and Borodin-Deift
\cite{BD} and coming up in point processes, is obtained by setting
$$
 \gamma_1^{\prime}=\al,~~
 \gamma_2^{\prime\prime}=\beta
~,~~\gamma=\gamma_2^{\prime}=
 \gamma_1^{\prime\prime}=0,~~ d_1=\xi,~ d_2=\xi^{-1},\mbox{all}~u_i=0,
~N_1=N_2=0.
 $$
 $$
  a=c=1,~~ b=-\xi-\xi^{-1}
 $$
in the weight (\ref{weight}). We have that
  $$ \al_i=(n+\al)\dt_{i1} ~~ \mbox{and}~~
 \beta_i=-(n+\beta)\dt_{i1}
  $$
  and so
$$
 \LR^{(n)}_1=(n+\al) L_1 ~~\mbox{and} ~~\LR^{(n)}_2=-(n+\beta)L_2.
 $$
Therefore (\ref{recurrence1}) and (\ref{recurrence2}) read
 $$
  \left( (n+\al+1) L_1-(n+\beta+1)L_2 \right)_{n+1,n+1}
 -
  \left( (n+\al-1) L_1-(n+\beta-1)L_2 \right)_{nn}
 =0
  $$

\bean
 \lefteqn{ (v_{n}(n+\al)L_1-(n+\beta)L_2)_{n+1,n}
  -(v_{n-1}(n-1+\al)L_1-(n-1+\beta)L_2)_{n,n-1}
  } \\
 && \hspace{9cm}+\bigl( L_1^2+b L_1\bigr)_{n,n}\\
 &=&
 \bigl(v_{1}(1+\al) L_1-(1+\beta)L_2\bigr)_{21}
  +\bigl( L_1^2+b L_1\bigr)_{11}
,\eean
  leading to a $3$-step relation and a $4$-step
  relation in $x_{n+1}$ and $y_{n+1}$,
  $$
 -(n+\al+1) \underline{\underline{x_{n+1}}}
 y_n +(n+\beta+1)\underline{\underline{y_{n+1}}}x_n
  +(n+\al-1)y_{n-1}x_n-(n+\beta-1)x_{n-1}y_n
 =0.$$
 and
 \bean
  \lefteqn{-v_n((n+\al+1)\underline{\underline{x_{n+1}}}y_{n-1} +n+\beta)
  +v_{n-1}((n+\al-2)x_{n}\underline{y_{n-2}} +n+\beta -1)}
  \\
&&\hspace{8cm}+x_n y_{n-1} (x_n y_{n-1} +\xi+\xi^{-1})
\\
 &=&-v_1
 \left(x_2(\al+2)+\beta+1\right)+x_1(x_1+\xi+\xi^{-1}).
 \eean
So, all $x_n$ and $y_n$ are rational expressions in
terms of $x_1,x_2,y_1,y_2$. Note these relations are
different from those found by Borodin \cite{B}.

\newpage

\section{Appendix 1: Virasoro algebras}

In \cite{AvM2}, we defined a Heisenberg and Virasoro
algebra of vector operators $^{\beta}\BJ^{(i)}_{k}$,
depending on a parameter $\beta
>0$:
$$ \left(
^{\beta}\BJ_k^{(1)}\right)_n=\,\,^{\beta}J_k^{(1)}+nJ_k^{(0)}
 ~\mbox{and}~ \left(\BJ_k^{(0)}\right)_n=nJ_k^{(0)}= n\dt_{0k}
,$$
 and
\vspace{0.3cm}

\noindent$\displaystyle{{}^{\beta}\BJ_k^{(2)}}$
\bea &=&{\beta} \sum_{i+j=k}
:{}^{\beta}\BJ_i^{(1)}{}^{\beta}\BJ_j^{(1)}:
+\left(1- {\beta} \right)\left((k+1) \,\,
^{\beta}\BJ_k^{(1)} -k~\BJ_k^{(0)}\right)\nonumber\\
&=&\left( {\beta} . ~{}^{\beta}J_k^{(2)} + \left(2n\beta +(k+1)(1-
{\beta} )\right).
 ~{}^{\beta}J_k^{(1)}
+  {n(n\beta+1-\beta)}  J_k^{(0)}\right)_{n\in
\BZ} .
 \nonumber\\
 \eea
The ${}^{\beta}\BJ_k^{(2)}$'s satisfy the commutation
relations: (see \cite{AvM2})
\begin{eqnarray}
\left[~
{}^{\beta}\BJ_k^{(1)},{}^{\beta}\BJ_{\ell}^{(1)}
  \right] &=&\frac{k}{2\beta}\delta_{k,-\ell}\nonumber\\
\left[~{}^{\beta}\BJ_k^{(2)},~{}^{\beta}\BJ_{\ell}^{(1)}
\right]
 &=&
  -\ell ~ ~{}^{\beta}\BJ_{k+\ell}^{(1)}
+\frac{k(k+1)}{2}\left(\frac{1}{\beta}-
1\right)\delta_{k,-\ell}\nonumber \\
\left[~{}^{\beta}\BJ_k^{(2)},~{}^{\beta}\BJ_{\ell}^{(2)}
 \right]
&=&(k-\ell)~{}^{\beta}\BJ_{k+\ell}^{(2)} +c\left(
\frac{k^3-k}{12} \right)\dt_{k,-\ell}~,
\end{eqnarray}
with central charge $$
c=1-6\left({\beta}^{1/2}
-{\beta}^{-1/2}  \right)^2. $$ In
the expressions above, \bea
 {}^{\beta}J_k^{(1)}&=&\frac{\pl}{\pl
t_k}~\mbox{for}~~k>0
 \nonumber\\
&=&\frac{1}{2\beta}(-k)t_{-k}~\mbox{for}~~k<0\nonumber\\
 &=&0~\mbox{for}~~k=0\nonumber\\  \nonumber\\
 ^{\beta}J^{(2)}_{k}&=&\sum_{i+j=k}\frac{\pl^2}{\pl
 t_{i}\pl t_{j}}+\frac{1}{\beta}\sum_{-i+j=k}it_{i}\frac{\pl}{\pl
 t_{j}}+\frac{1}{4\beta^2}\sum_{-i-j=k}it_{i}jt_{j}
 \eea
In particular, for $\beta = 1/2$ and $1$, the
 ${}^{\beta}\BJ_k^{(2)}$ take on the form:
\bea {}^{\beta}\BJ_k^{(2)}(t)\Big|_{\beta=1/2}
 %=\left(J_{k,n}^{(2)}\right)_{n\in\BZ}
 &=&\left.\frac{1}{2}\left(~{}^{\beta}J_k^{(2)}
+(2n+k+1)\,\, ^{\beta}J_k^{(1)} +n(n+1)J_k^{(0)}
 \right)_{n\in\BZ}\right|_{\beta=1/2},\nonumber\\   &&\\
 {}^{\beta}\BJ_k^{(2)}(t)\Big|_{\beta=1}
 %=\left(J_{k,n}^{(2)}\right)_{n\in\BZ}
 &=&\left.\left(~{}^{\beta}J_k^{(2)}
+2n\,\, ^{\beta}J_k^{(1)} +n^2 J_k^{(0)}
 \right)_{n\in\BZ}\right|_{\beta=1}
\eea

%\newpage

\section{Appendix 2: Useful formulae about the
 \\ Toeplitz lattice}

The first Toeplitz vector field
(\ref{Toeplitz}), corresponding to the
Hamiltonians
 $$
 H_1^{(1)}=-\Tr~ L_1=\sum_0^{\iy} x_{i+1}y_i,~
 H_1^{(2)}=-\Tr~ L_2=\sum_0^{\iy} x_{i}y_{i+1},
 $$
 and
 \bean
  H_2^{(1)}&=&-\frac{1}{2}\Tr~ L_1^2=
 -\frac{1}{2}\sum_0^{\iy} x_{i+1}^2y_{i}^2+
  \sum_0^{\iy} y_ix_{i+2} v_{i+1}\\
 H_2^{(2)}&=&-\frac{1}{2}\Tr~ L_2^2=
 -\frac{1}{2}\sum_0^{\iy} x_i^2y_{i+1}^2+
  \sum_0^{\iy} x_iy_{i+2} v_{i+1}
  \eean
 reads: \bean
\frac{\pl x_n}{\pl t_1}=v_nx_{n+1}  &~~~~~& \frac{\pl
y_n}{\pl t_1}=-v_ny_{n-1}   \\ \frac{\pl x_n}{\pl
s_1}=v_nx_{n-1}   &~~~~~& \frac{\pl y_n}{\pl
s_1}=-v_ny_{n+1}  . \\ \eean
and
{\footnotesize
\bean \frac{\pl x_n}{\pl t_2}&=&
  v_{n}\frac{\pl H^{(1)}_2}{\pl y_n}
  =
  - v_{n}
  \Bigl(x_{n+1}^2y_n-x_{n+2}v_{n+1}
 +
  x_nx_{n+1}y_{n-1}\Bigr)
  \\&&\\
   \frac{\pl y_n}{\pl t_2}&=&
  -v_{n}\frac{\pl H^{(1)}_2}{\pl x_n}
  = v_{n}
  \Bigl(x_{n}y_{n-1}^2-y_{n-2}v_{n-1}+
  y_ny_{n-1}x_{n+1}\Bigr)
  \\&&\\
  \frac{\pl x_n}{\pl s_2}&=&
  v_{n}\frac{\pl H^{(2)}_2}{\pl y_n}
  =
  - v_{n}
  \Bigl(x_{n-1}^2y_n-x_{n-2}v_{n-1}
 +
  x_nx_{n-1}y_{n+1}\Bigr)
 \\&&\\
 \frac{\pl y_n}{\pl s_2}&=&
  -v_{n}\frac{\pl H^{(2)}_2}{\pl x_n}
  = v_{n}
  \Bigl(x_{n}y_{n+1}^2-y_{n+2}v_{n+1}+
  y_ny_{n+1}x_{n-1}\Bigr)
\eean

}

%\newpage

\section{Appendix 3: Proof of Theorem 0.2}

Before giving the proof of Theorem 0.2, we need

\begin{lemma}

\bean
 \left(L_1^{N+1}\right)_{nn}&=&
 -x_{n+N}y_{n-1}\prod_1^{N}v_{n+i-1} +\ldots
 -x_ny_{n-N-1}\prod_1^N v_{n-i}\\&&\\
\left(L_2^{N+1}\right)_{nn}&=&
 -y_{n+N}x_{n-1}\prod_1^{N}v_{n+i-1} +\ldots
  -y_nx_{n-N-1}\prod_1^N v_{n-i}\\&&\\
v_n\left(L_1^{N+1}\right)_{n+1,n}&=&
 -x_{n+N+1}y_{n-1}\prod_0^{N}v_{n+i} +\ldots
  -x_{n+1}y_{n-N-1}\prod_0^N v_{n-i}
  \\&&\\
\left(L_2^{N+1}\right)_{n+1,n}&=&
 -y_{n+N}x_{n}\prod_0^{N-1}v_{n+i} +\ldots
 -x_{n-N}y_{n}\prod_1^N v_{n-i+1}.%\\&&\\
\eean
The two highest and two lowest terms in $$
  \LR_1^{(n)}=\sum_{1}^{N_1+1} \al_i(t)  L_1^i
   ~~~\mbox{and}~~\LR_2^{(n)} =-\sum_{1}^{N_2+1}\beta_i (t)
   L_2^i,
   $$
 are the following:
{\footnotesize \bean
 (\LR_1^{(n)})_{nn}&=&
 -\al_{N_1+1}
 \left( x_{n+N_1}y_{n-1}\prod_1^{N_1}v_{n+i-1} +\ldots
 +x_ny_{n-N_1-1}\prod_1^{N_1} v_{n-i}\right)\\
 &&
 -\al_{N_1}
 \left( x_{n+N_1-1}y_{n-1}\prod_1^{N_1-1}v_{n+i-1} +\ldots
 +x_ny_{n-N_1}\prod_1^{N_1-1} v_{n-i}\right)~+\ldots
 % v_n \ldots
% v_{n-1+N_1}~y_{n-1}~\underline{x_{n+N_1}}~+\ldots
 \\
 (\LR_2^{(n)})_{nn}&=&
 \beta_{N_2+1}\left(y_{n+N_2}x_{n-1}\prod_1^{N_2}
  v_{n+i-1} +\ldots
  +y_nx_{n-N_2-1}\prod_1^{N_2} v_{n-i}
  \right)\\
 &&  +\beta_{N_2}\left(y_{n+N_2-1}x_{n-1}\prod_1^{N_2-1}
  v_{n+i-1} +\ldots
  +y_nx_{n-N_2}\prod_1^{N_2-1} v_{n-i}
  \right)+\ldots
 \\&&\\
 v_{n}( \LR_1^{(n)})_{n+1,n}
 &=&
 -\al_{N_1+1}\left( x_{n+N_1+1}y_{n-1}\prod_0^{N_1}
 v_{n+i}
   +\ldots
  +x_{n+1}y_{n-N_1-1} \prod_0^{N_1} v_{n-i}\right)\\
  &&
   -\al_{N_1}\left( x_{n+N_1}y_{n-1}\prod_0^{N_1-1}
 v_{n+i}
   +\ldots
  +x_{n+1}y_{n-N_1} \prod_0^{N_1-1} v_{n-i}\right)~+\ldots
  \\
 ( \LR_2^{(n)})_{n+1,n}
 &=&
 \beta_{N_2+1} \left(y_{n+N_2}x_{n}\prod_0^{N_2-1}
 v_{n+i} +\ldots
 +x_{n-N_2}y_{n}\prod_1^{N_2} v_{n-i+1}\right)  \\
 &&
  +\beta_{N_2} \left(y_{n+N_2-1}x_{n}\prod_0^{N_2-2}
 v_{n+i} +\ldots
 +x_{n-N_2+1}y_{n}\prod_1^{N_2-1} v_{n-i+1}\right)  ~+\ldots
  \eean
  }

with

\bean \al_{N_1+1}&=&c\left(u_{N_1}+\dt_{N_1,0}
(n+\gamma_1^{\prime}+\gamma_2^{\prime}+\gamma)\right)\\
      \al_{N_1}&=&c\left(u_{N_1-1}+\dt_{N_1,1}
(n+\gamma_1^{\prime}+\gamma_2^{\prime}+\gamma)\right)
 +bu_{N_1}\\
      -\beta_{N_2+1}&=&a
       \left(u_{-N_2}+\dt_{N_2,0}
(n+\gamma_1^{\prime\prime}+\gamma_2^{\prime\prime}-
\gamma)\right)
 \\
      -\beta_{N_2}&=&a\left(u_{-N_2+1}+\dt_{N_2,1}
(n+\gamma_1^{\prime\prime}+\gamma_2^{\prime\prime}-
\gamma)\right)
 +bu_{-N_2}\\
      \eean

\end{lemma}

%\newpage

{\medskip\noindent{\it Proof of Theorem
0.2:\/} }

\noindent CASE 1: $a,c \neq 0$:
 Equations (\ref{recurrence1}) and (\ref{recurrence2})
 are two inductive equations, one having
 $(N_1+N_2+4)$ steps and the other having $(N_1+N_2+3)$
  steps. In the equations below, we underline twice
  the highest terms and once the lowest ones.
  The exact equations (\ref{recurrence1})
  and (\ref{recurrence2}) are denoted by
  (\ref{recurrence1})${}_{n}$
  and (\ref{recurrence2})${}_{n}$.

\bigbreak

\noindent $\bullet ~~N_1=N_2=N$:
 The two equations
 (\ref{recurrence1})${}_{n+1}$
  and (\ref{recurrence2})${}_{n}$, form a system of
  two equations in the unknowns
  $x_{n+N+2}$ and $y_{n+N+2}$, the variable with the lowest index
   being $y_{n-N-1}$ :
{\footnotesize
\bean
 \mbox{(\ref{recurrence1})${}_{n+1}$}
 &=&
  -\left(\al_{N+1}
  \underline{\underline{x_{n+N+2}}}~y_{n+1}
  +\beta_{N+1}\underline{\underline{y_{n+N+2}}}~x_{n+1}
  \right)\prod_1^{N}v_{n+i+1} +\ldots
\\
  \mbox{(\ref{recurrence2})${}_{n}$}
 &=&
  -\al_{N+1}\underline{\underline{x_{n+N+2}}}~y_{n}
  \prod_1^{N}v_{n+i+1}+\ldots
%  \\
%  && \ldots
+\al_{N+1}\underline{y_{n-N-1}}~x_{n+1}
  \prod_0^{N}v_{n-i}
 \eean
}

\noindent $\bullet ~~N_1=N,~~N_2=N+1$:
 The two equations (\ref{recurrence1})${}_{n}$
  and (\ref{recurrence2})${}_{n}$ form a system of
  two equations in the unknowns
  $x_{n+N+2}$ and $y_{n+N+2}$, the variable with the lowest index
   being $x_{n-N-2}$ :

\bean \mbox{(\ref{recurrence1})${}_{n}$}
 &=&
%  -\al_{N+1}\underline{\underline{x_{n+N+2}}}y_{n+1}
%  \prod_1^{N}v_{n+i+1}
  -\beta_{N+2}\underline{\underline{y_{n+N+2}}}x_{n}
  \prod_1^{N+1}v_{n+i} +\ldots+
  \beta_{N+2} y_n \underline{x_{n-N-2}}
   \prod_1^{N+1}v_{n-i}
\\
  \mbox{(\ref{recurrence2})${}_{n}$}
 &=&
  -\left(
   \al_{N+1}\underline{\underline{x_{n+N+2}}}y_{n}
 +\beta_{N+2} \underline{\underline{
y_{n+N+2}}}x_{n+1}\right)\prod_0^{N}v_{n+i+1}+\ldots
 \eean

\noindent $\bullet ~~N_1=N,~~N_2=N-1$
:  The two equations (\ref{recurrence1})${}_{n}$
  and the dual equation
   (\ref{recurrence2})$\tilde{}_{n}$,
   via the involution $\tilde{}$, form a system of
  two equations in the unknowns
  $x_{n+N+1}$ and $y_{n+N+1}$, the variable with the lowest index
   being $y_{n-N-1}$ :

\bean
 \mbox{(\ref{recurrence1})${}_{n}$}
 &=&
  -\al_{N+1}\underline{\underline{x_{n+N+1}}}y_{n}
  \prod_1^{N}v_{n+i}+\ldots
%  \\
%  && \ldots
+\al_{N+1}\underline{y_{n-N-1}}x_{n}
  \prod_1^{N}v_{n-i} \\
  \mbox{(\ref{recurrence2})${}^{\top}_{n}$}
 &=&
  -\left(\al_{N+1}
   \underline{\underline{x_{n+N+1}}}y_{n+1}
  +\beta_{N}\underline{\underline{y_{n+N+1}}}x_{n}
 \right) \prod_0^{N-1}v_{n+i+1} +\ldots
 \eean

%\newpage

\noindent CASE 2:~All cases below lead to two inductive
 equations, one having
 $(N_1+N_2+3)$ steps and the other having $(N_1+N_2+2)$
  steps:

  \medbreak

\noindent $\bullet ~~N_1=N,~~N_2=N$, $a=0,~b=c=1$:
  The two equations
 (\ref{recurrence1})${}_{n}$
  and (\ref{recurrence2})$\tilde{}_{n}$, form a system of
  two equations in the unknowns
  $x_{n+N+1}$ and $y_{n+N+1}$, the variable with the lowest index
   being $y_{n-N-1}$ :
\bean
 \mbox{(\ref{recurrence1})${}_{n}$}
 &=&
  -\al_{N+1} \underline{\underline{x_{n+N+1}}}y_{n}
  \prod_1^{N}v_{n+i}+\ldots
%  \\
%  && \ldots
 +\al_{N+1} \underline{y_{n-N-1}}x_{n}
  \prod_1^{N}v_{n-i}
  \\
   \mbox{(\ref{recurrence2})$\tilde{}_{n}$}
 &=&
  -\left(\al_{N+1}
   \underline{\underline{x_{n+N+1}}}y_{n+1}
  +\beta_{N}\underline{\underline{y_{n+N+1}}}x_{n}
 \right) \prod_0^{N-1}v_{n+i+1} +\ldots
 \eean

\noindent $\bullet ~~N_1=N,~~N_2=N+1$, $a=0,~b=c=1$:
 The two equations (\ref{recurrence1})${}_{n+1}$
  and (\ref{recurrence2})${}_{n}$ form a system of
  two equations in the unknowns
  $x_{n+N+2}$ and $y_{n+N+2}$, the variable with the lowest index
   being $y_{n-N-1}$ :
\bean \mbox{(\ref{recurrence1})${}_{n+1}$}
 &=&
  -\left(\al_{N+1}
  \underline{\underline{x_{n+N+2}}}y_{n+1}
  +\beta_{N+1}\underline{\underline{y_{n+N+2}}}x_{n+1}
  \right)\prod_1^{N}v_{n+i+1} +\ldots
\\
  \mbox{(\ref{recurrence2})${}_{n}$}
 &=&
  -\al_{N+1}\underline{\underline{x_{n+N+2}}}y_{n}
  \prod_0^{N}v_{n+i+1}+\ldots
%  \\
%  && \ldots
+\al_{N+1}\underline{y_{n-N-1}}x_{n+1}
  \prod_0^{N}v_{n-i}
 \eean

 \noindent $\bullet ~~ N_1=N,~~N_2=N-1$
  $c=0,~a=b=1$:  The two equations
    (\ref{recurrence1})${}^{}_{n+1}$
  and the equation (\ref{recurrence2})$\tilde{}_{n}$ form a system of
  two equations in the unknowns
  $x_{n+N+1}$ and $y_{n+N+1}$, the variable with the lowest index
   being $x_{n-N}$ :
\bean
  \mbox{(\ref{recurrence1})${}_{n+1}$}
 &=&
  -\left(\al_{N}
   \underline{\underline{x_{n+N+1}}}y_{n+1}
  +\beta_{N}\underline{\underline{y_{n+N+1}}}x_{n+1}
 \right) \prod_1^{N-1}v_{n+i+1} +\ldots \\
 \mbox{(\ref{recurrence2})$\tilde{}_{n}$}
 &=&
  -\beta_{N}\underline{\underline{y_{n+N+1}}}x_{n}
  \prod_0^{N-1}v_{n+i+1}+\ldots
%  \\
%  && \ldots
+\beta_{N}\underline{x_{n-N}}y_{n+1}
  \prod_0^{N-1}v_{n-i} \\
 \eean

CASE 3. All cases below lead to two inductive
equations, both having $N_1+N_2+1$ steps. Using again
Lemma 7.1, one searches for the highest and lowest terms
 in the relations (\ref{recurrence3}):
\bean
 0&=&  nx_n+\displaystyle{\frac{v_n}{y_n}}
  \left(
  \begin{array}{l}
  -\left( L_1 P_1^{\prime} (
L_1)\right)_{n+1,n+1}
  - \left(  L_2 P_2^{\prime} (
L_2)\right)_{n,n} \\  \\
 + (P_1^{\prime}( L_1))_{n+1,n}
  + (P_2^{\prime}( L_2))_{n,n+1}
 \\
 \end{array}\right)
 \\
 &=&
 \frac{v_n}{y_n}
 \left(
  \begin{array}{l}
   -u_{N_1}
(L_1^{N_1})_{n+1,n+1}
  -u_{-N_2} (L_2^{N_2})_{nn}+\dots \\ \\
 + u_{N_1} (L_1^{N_1-1})_{n+1,n}
 + u_{-N_2} (L_2^{N_2-1})_{n,n+1}+\dots
 \end{array} \right)
 \\&=&
 u_{N_1}\frac{v_n}{y_n}\left(x_{n+N_1}y_{n} \prod_1^{N_1-1}v_{n+i}+\ldots
 +x_{n+1}y_{n-N_1+1} \prod_1^{N_1-1}v_{n+1-i}\right)
 \\&&
 -u_{N_1}\frac{v_n}{y_n}\left(\ldots+
 x_{n+1}y_{n-N_1+1}\prod_1^{N_1-2}v_{n-i}
 \right)
 \\&&
 + u_{-N_2}\frac{v_n}{y_n}\left(y_{n+N_2-1}x_{n-1} \prod_1^{N_2-1}
 v_{n+i-1}+\ldots
 +y_{n}x_{n-N_2} \prod_1^{N_2-1}v_{n-i}\right)
 \\&&
 -u_{-N_2}\frac{v_n}{y_n}\left(y_{n+N_2-1}x_{n-1} \prod_1^{N_2-2}
 v_{n+i}+\ldots
 \right)+\ldots
 \\&=&
 u_{N_1}\left(x_{n+N_1}  \prod_0^{N_1-1}v_{n+i}+\ldots
 -x_{n+1}y_{n-N_1+1}x_n  \prod_0^{N_1-2}v_{n-i}\right)
 \\&&
 -u_{-N_2}\left(y_{n+N_2-1}x_{n-1}x_n \prod_0^{N_2-2}
 v_{n+i}+\ldots
 -x_{n-N_2} \prod_0^{N_2-1}v_{n-i}\right)+\ldots,
\eean
 and, by duality,
\bean
 0&=&  ny_n+ \displaystyle{\frac{v_n}{x_n}  }
\left(
\begin{array}{l}
  -\left( L_1 P_1^{\prime} (
L_1)\right)_{n,n}
 -\left(  L_2 P_2^{\prime} (
L_2)\right)_{n+1,n+1} \\  \\
 + (P_1^{\prime}(
L_1))_{n+1,n}
  + (P_2^{\prime}( L_2))_{n,n+1}
  \end{array}\right) \\
  \\&=&
 -u_{N_1} \left(x_{n+N_1-1}y_{n-1}
   y_n \prod_0^{N_1-2}v_{n+i}+\ldots
 - y_{n-N_1}  \prod_0^{N_1-1}v_{n-i}\right)
 \\&&
 +u_{-N_2} \left(y_{n+N_2} \prod_0^{N_2-1}
 v_{n+i}+\ldots
 -y_{n+1}x_{n-N_2+1} y_n\prod_0^{N_2-2}v_{n-i}\right)
+\ldots. \eean
Here again, one uses different indices $n$ for each of
the cases:
%
%\newpage

\noindent $\bullet ~~ N_1=N,~~N_2=N$
  \bean
\mbox{(\ref{recurrence3})}_n&=&u_{N}
 \underline{\underline{x_{n+N}}}
\prod_0^{N-1}v_{n+i}+\ldots
+u_{-N} \underline{x_{n-N}} \prod_0^{N-1}v_{n-i}
 +\ldots\\
 \mbox{(\ref{recurrence3})}\tilde{}_n&=&
 u_{-N}\underline{\underline{y_{n+N}}}
\prod_0^{N-1}v_{n+i}+\ldots
+u_{N} \underline{y_{n-N}} \prod_0^{N-1}v_{n-i}
 +\ldots
  \eean

\noindent $\bullet ~~ N_1=N,~~N_2=N+1$
 \bean
\mbox{(\ref{recurrence3})}_{n+1}&=& \left(u_{N}
 \underline{\underline{x_{n+N+1}}}
  -u_{-N-1}
 \underline{\underline{y_{n+N+1}}}x_{n}x_{n+1}
 \right)%y_{n+1}
 \prod_0^{N-1}v_{n+i+1}+\ldots
  \\&&+u_{-N-1}%y_{n+1}
   \underline{ x_{n-N}}
   \prod_0^{N}v_{n-i+1}+\ldots
 \\
 \mbox{(\ref{recurrence3})}\tilde{}_n&=&
 u_{-N-1}\underline{\underline{y_{n+N+1}}}%x_{n}
\prod_0^{N}v_{n+i}+\ldots
\\&&
 -\left(u_{-N-1}y_{n+1}\underline{x_{n-N}} y_n
 -u_{N} \underline{ y_{n-N}}
  \right)%x_n
  \prod_0^{N-1}v_{n-i}  +\ldots
 \eean

\noindent $\bullet ~~N_1=N,~~N_2=N-1$,%@@@@@@@@@
 \bean
\mbox{(\ref{recurrence3})}_{n}&=&
u_{N}\underline{\underline{x_{n+N}}}%y_{n}
\prod_0^{N-1}v_{n+i}+\ldots
\\&&
 -\left(u_{N}x_{n+1}\underline{y_{n-N+1}}x_n
 -u_{-N+1} \underline{ x_{n-N+1}}
  \right)%y_n
   \prod_0^{N}v_{n-i}+\ldots
 \\
\mbox{(\ref{recurrence3})}{\tilde {}}_{n+1}&=& \left(u_{-N+1}
 \underline{\underline{y_{n+N}}}
  -u_{N}
 \underline{\underline{x_{n+N}}}y_{n}y_{n+1}
 \right)%x_{n+1}
  \prod_0^{N-2}v_{n+i+1}+\ldots
  \\&&+u_{N}%x_{n+1}
  \underline{ y_{n-N+1}}
   \prod_0^{N-1}v_{n-i+1} +\ldots
 \eean
This ends the proof of Theorem 0.2. \qed

\end{document}